\documentclass[onefignum,onetabnum]{siamonline190516}

\usepackage{lipsum}
\usepackage{amsfonts}
\usepackage{amssymb}
\usepackage{graphicx}
\usepackage{subfig}
\usepackage{epstopdf}
\usepackage{algorithmic}
\usepackage{bbm}
\usepackage{bm}
\ifpdf
  \DeclareGraphicsExtensions{.eps,.pdf,.png,.jpg}
\else
  \DeclareGraphicsExtensions{.eps}
\fi

\usepackage{enumitem}
\setlist[enumerate]{leftmargin=.5in}
\setlist[itemize]{leftmargin=.5in}

\newsiamremark{remark}{Remark}
\newsiamremark{hypothesis}{Hypothesis}
\crefname{hypothesis}{Hypothesis}{Hypotheses}
\newsiamthm{claim}{Claim}

\usepackage{pgf}
\usepackage{tikz}
\usetikzlibrary{matrix}
\usetikzlibrary{decorations.pathmorphing} 
\usetikzlibrary{fit}					
\usetikzlibrary{backgrounds}	
\usetikzlibrary{shapes.arrows}
\usetikzlibrary{shapes.geometric}
\usetikzlibrary{shapes.multipart}
\usetikzlibrary{matrix,arrows,decorations.pathreplacing,positioning}
\tikzstyle{rec}=[draw,rectangle, minimum height=2cm]
\tikzset{>=stealth', punkt/.style={rectangle, 
fill=gray!40, 
draw=black, very thick, text width=3em, minimum height=2.5em, text centered}}
\tikzset{>=stealth', punkt_l/.style={rectangle, 
fill=gray!40, 
draw=black, very thick, text width=4em, minimum height=2.5em, text centered}}
\tikzstyle{background}=[rectangle,fill=green!20,inner sep=0.1cm,rounded corners=4mm,minimum height=3em]
\tikzstyle{sum} = [draw, fill=gray!40, circle, node distance=1cm]
\tikzstyle{dot} = [circle, fill=black, inner sep=0pt, minimum size=5pt, node contents={}]

\def\*#1{\mathbf{#1}}
\def\+#1{\mathcal{#1}}
\def\-#1{\mathbb{#1}}

\def\Real{\mathbb{R}}
\def\Expect{\mathbb{E}}

\def\Order{\mathcal{O}}

\renewcommand{\tilde}{\widetilde}

\headers{Compressive Computed Tomography Reconstruction through Denoising AMP}{A. Perelli, M. Lexa, A. Can, and Mike E. Davies}

\title{Compressive Computed Tomography Reconstruction through Denoising Approximate Message Passing\thanks{Submitted to the editors of SIIMS.
\funding{This work was supported by U.S. Department of Homeland Security, Science and Technology Directorate, Explosives Division, BAA 13-05, Contract \# HSHQDC-14-C-B0048, EPSRC platform grant EP/J015180/1 and the ERC Advanced grant, project C-SENSE, (ERC-ADG-2015-694888). A. Perelli has received funding from the European Union's Horizon 2020 research and innovation programme under the Marie Sklodowska-Curie grant agreement no. 713683 (COFUNDfellowsDTU).}}}

\author{Alessandro Perelli\thanks{Technical University of Denmark, DTU, Lyngby, DK 
  (\email{alper@dtu.dk}).}
\and Michael Lexa\thanks{GE Research Center, Niskayuna, US
(\email{lexa@ge.com}, \email{can@ge.com}).}
\and Ali Can\footnotemark[3]
\and Mike E. Davies\thanks{Institute for Digital Communications (IDCOM), The University of Edinburgh, UK (\email{m.davies@ed.ac.uk})}}

\usepackage{amsopn}

\ifpdf
\hypersetup{
  pdftitle={Compressive Computed Tomography Reconstruction through Denoising Approximate Message Passing},
  pdfauthor={A. Perelli, M.  Lexa, A. Can, and M. Davies}
}
\fi

\begin{document}

\maketitle

\begin{abstract}
X-ray Computed Tomography (CT) reconstruction from a sparse number of views is a useful way to reduce either the radiation dose or the acquisition time, for example in fixed-gantry CT systems, however this results in an ill-posed inverse problem whose solution is typically computationally demanding. Approximate Message Passing (AMP) techniques represent the state of the art for solving undersampling Compressed Sensing problems with random linear measurements but there are still not clear solutions on how AMP should be modified and how it performs with real world problems. This paper investigates the question of whether we can employ an AMP framework for real sparse view CT imaging? The proposed algorithm for approximate inference in tomographic reconstruction incorporates a number of advances from within the AMP community, resulting in the Denoising Generalised Approximate Message Passing CT algorithm (D-GAMP-CT). Specifically, this exploits the use of sophisticated image denoisers to regularise the reconstruction. While in order to reduce the probability of divergence the (Radon) system and Poission non-linear noise model are treated separately, exploiting the existence of efficient preconditioners for the former and the generalised noise modelling in GAMP for the latter. Experiments with simulated and real CT baggage scans confirm that the performance of the proposed algorithm outperforms statistical CT optimisation solvers. 
\end{abstract}



\section{Introduction}

X-ray Computed Tomography (CT) is one of the most widely used imaging techniques for medical diagnosis, image-guided radiotherapy, material characterization and security applications. Reducing X-ray radiation exposure is an important concern in particular for diagnostic CT where patients are subjected to repeated scans and for CT baggage scanners since the transmitted energy is related to the lifetime of the X-ray source. Furthermore, CT scanners employing Dual Energy (DE) systems tend to either reduce the acquisition data per energy or increase the dose and acquisition time. To lower the X-ray dose, two different strategies can be implemented: reducing the X-ray flux toward each detector element, i.e. the milliampere per seconds (low-mAs) per projection, or decrease the number of projections (sparse-views) per rotation. Similarly fixed gantry systems, e.g. \cite{gonzales2014},  designed to accelerate scan time tend to further restrict the set of projections that can be acquired.

CT image reconstruction from sparse views and low dose, achieved by conventional filtered back projection (FBP) algorithms, is generally affected by noticeable streaking artifacts, due to insufficient sampling, and is not of acceptable quality for diagnostic purposes \cite{bruyant2000}. There is therefore a need in CT imaging applications for high quality image reconstruction algorithms that can accommodate sparse views and low dose. Many approaches have been proposed to solve this problem \cite{sidky2006}. In particular, state of the art statistical image reconstruction typically aims to minimize a cost function defined as a sum of a data fidelity term that takes into account the measurement's statistical model and the geometry of the acquisition system, and a regularization term that imposes a prior model on the solution. Generally, the cost function for X-ray CT is either the negative log-likelihood function \cite{elbakri2002} or a penalized weighted least-squares (PWLS) cost function with a weighted quadratic approximation of the Poisson measurement noise model \cite{huang2013}, \cite{niu2014}. Although several types of iterative algorithms have been designed to solve the statistical X-ray CT problem which can provide images with enhanced resolution and reduced artifacts compared to the FBP \cite{sauer1993}, in general current methods require many iterations to converge yielding a high computation time, and are often not suitable for clinical/industrial CT uses \cite{beister2012}. 

A large number of iterative algorithms have been utilized for statistical CT reconstruction, among these are coordinate descent \cite{yu2011}, preconditioned conjugate gradient \cite{fessler1999} and ordered subsets \cite{erdogan1999}. Recently researchers have developed new algorithms with faster convergence by using splitting techniques \cite{ramani2012}, alternating direction method of multipliers based algorithm \cite{chun2014} or combining Nesterov momentum techniques with ordered subsets to accelerate gradient descent methods \cite{kim2015}. 
In general, any first-order iterative method requires at each iteration the computation of at least one forward and back projection operator, together with a proximal mapping to account for the regularization term. These represent the main contributions to the overall computational time. In order to accelerate the reconstruction, it is therefore necessary to either design faster CT operators or develop iterative algorithms that can converge in fewer iterations. 

In this work, we investigate the use of an emerging reconstruction method from Compressed Sensing (CS), called Approximate Message Passing (AMP) \cite{donoho2009}, for sparse view CT reconstruction. AMP based inference refers to a family of iterative algorithms first proposed in \cite{donoho2009} for Compressed Sensing problems with an i.i.d. random Gaussian system matrix and a sparse signal model. AMP is a form of approximate Bayesian inference based on the notion of message passing or loopy belief propagation and is also strongly connected to the family of Expectation Propagation and Expectation Consistent approximation algorithms \cite{opper2005}. In essence, message passing algorithms work by iteratively updating marginal probabilities on the unknown variables until a locally consistent posterior probability model is obtained. The compelling aspect of the AMP family of algorithms is that they are designed to work in the large system limit (for random systems) which enables the central limit theorem to be invoked. This in turn simplifies the messages to be Gaussian distributions, requiring the algorithm to only pass means and variances. The result is a very efficient algorithm that is remarkably similar to the more traditional iterative shrinkage algorithm but with an additional "Onsager correction term" \cite{donoho2009}. It  also has many similarities to the Alternating Direction Method of Multipliers (ADMM) algorithm \cite{rangan2015}. 

\newpage 

Today, AMP based algorithms provide the state-of-the-art performance in CS reconstruction both in terms of computation and reconstruction performance, e.g. \cite{vila2013,guo2015,metzler2014}. For Gaussian measurements and random i.i.d. Gaussian sensing matrix, the convergence of the algorithm has been theoretically proven and, furthermore, it 
can be accurately quantified through its state evolution (SE) equations, exhibiting exponential convergence, in practice converging in very few iterations, for soft-thresholding function \cite{bayati2011lasso}, linear denoisers \cite{bayati2011} and a class of non-separable denoisers \cite{berthier2020}. AMP can also incorporate non-Gaussian noise models through Rangan's Generalized Approximate Message Passing (GAMP) \cite{rangan2011} and can approximate the Minimum Mean Square Error (MMSE) estimator by using a correctly matched prior or by exploiting learning structures such as Expectation Maximization \cite{vila2013} or SURE \cite{guo2015}. It has even been shown to be capable of incorporating sophisticated black box denoising algorithms in place of a signal prior model, resulting in the Denoising AMP (D-AMP) framework \cite{metzler2014}.
However, a key criticism directed at AMP, and its generalizations, is that they are specialist algorithms for i.i.d. and related measurement matrices and hence it is unclear to what extent they can be successfully applied to real word sensing problems. 

There has been some work exploring the convergence properties of AMP and its generalizations to other matrix classes \cite{rangan2014}, S-AMP \cite{cakmak2014}, and linking the algorithm with more classical optimization strategies such as ADMM \cite{rangan2015}. A key problem of AMP is that when the measurement matrix is poorly conditioned and/or contains a significant mean offset the algorithm tends to diverge. One strategy for tackling this that is commonly used in loopy belief propagation is to incorporate damping to help stabilize the algorithm, \cite{schniter2015,vila2015}. However, damping comes at the cost of significantly reducing the algorithm's convergence speed. It is also not clear what the value of the Onsager term is for general (deterministic) measurement matrices and whether the SE equations still provide a good prediction of the algorithm's performance. 

Finally, Vector AMP \cite{rangan2016} is a class of convergent algorithms with poorly conditioned matrices but the behavior with Fourier matrices is not rigorously understood. In summary, VAMP is a promising method for Fourier-based imaging while we do not know whether AMP based techniques can provide a competitive reconstruction framework to state-of-the-art methods for general real world imaging problems. The aim of this paper is to explore these issues for the specific case of sparse view CT imaging.

\subsection{Main Contributions}
Our approach to develop an AMP based algorithm for CT reconstruction builds on a number of the recent developments in the field and, in particular, it makes use of the following key points: i) the design of a good preconditioner for the system based on the forward measurement model; ii) the inclusion of a non-linear Poisson noise model through the GAMP formulation; and iii) the incorporation of a broader class of signal prior than sparsity based models, through the D-AMP framework to enable the exploitation of state-of-the-art image denoising functions. We also demonstrate empirically the value of the Onsager term in the resulting algorithm and the accuracy of the generalized state evolution equations \cite{rangan2011} even in this non-random setting.

As far as the authors are aware, this is the first work aimed at designing a denoising message passing based algorithm for CT reconstruction. A key challenge in applying GAMP to CT is the fact that the CT measurement operator for parallel or fan beam geometry has the form of a Radon type transform and is very ill-conditioned. This would require a significant level of damping to stabilize it and would be extremely slow \cite{rangan2014}. The solution that we follow here is to replace the ill-conditioned operator with a much better conditioned one through preconditioning, exploiting the filtered back projection property of the system model \cite{nilchian2013}. The same procedure can be applied for different CT geometries like 2D fan-beam and 3D helical. 

Another key challenge for CT reconstruction is how to accurately represent the Poisson noise model in the system. This can be approximated as a weighted $L_2$ error criterion \cite{erdogan1999}, but then the preconditioner needs to account for both the system operator and the weighting matrix. While such preconditioners have been proposed, e.g. \cite{kim2015}, they do not exploit the geometry of the measurement system and the subsequent system remains poorly conditioned, resulting  in only modest improvements in convergence. In contrast, we will see that in the GAMP framework \cite{rangan2011} the system operator and the noise process are naturally decoupled. This allows us a fully exploit a geometric preconditioner \cite{nilchian2013}.

The final ingredient of our algorithm, which we call D-GAMP-CT, is the incorporation of the Block-matching and 3D filtering (BM3D) denoiser \cite{dabov2007} to implicitly define a signal model through a sophisticated denoiser, rather than simply a sparse factorizable prior distribution \cite{metzler2014}. While the proposed approach can leverage generic denoisers, we have utilized the BM3D denoiser since it provides state-of-the-art accuracy performance among deterministic denoisers and it also exploits a new implementation with reduced computational complexity \cite{perelli2015compressive}. 
We note that this framework is not restricted to the use of the BM3D denoiser but can be further extended to deep learning-based denoisers \cite{deepCNN2017}. However for deep learning-based denoisers, it is crucial for achieving high denoising performance to have a high quality noiseless training database and it is often challenging or infeasible to obtain noiseless images in medical imaging.

We will see that the flexibility of using such a denoiser within GAMP yields to a better reconstruction of the image structure compared to more popular regularization, such as Total Variation (TV) minimization. 

\subsection{Relation to Existing Work}\label{subsec:rel_works}

The main issue of stabilizing AMP algorithms for non i.i.d. measurement matrices has already received attention in the literature. As previously discussed, damping is a popular solution \cite{rangan2014,schniter2015,vila2015} and, for example, has been applied successfully to hyperspectral imaging reconstruction \cite{vila2014empirical,Baron2016}. Schemes have also been proposed for modifying the algorithm when the matrix contains a significant non-zero offset \cite{vila2015,manoel2015}. These approaches are fundamentally different from the one we present here where both issues are solved through our choice of a geometric preconditioner. Other aspects of our algorithm, such as the exploitation of general denoisers \cite{metzler2014,berthier2020}, and the use of generalized noise models \cite{rangan2011} have already appeared in the literature. Here we combine these to define a state of the art algorithm for sparse view CT reconstruction.

A new class of AMP algorithms called Vector AMP (VAMP) \cite{rangan2016} (and the similar orthogonal AMP in \cite{ma2017}) that directly tackle the ill-conditioning problem in AMP by exploiting the singular value decomposition (SVD) of the measurement matrix. Such algorithms exhibit impressive performance and have provable reconstruction guarantees for the class of right-orthogonally invariant random matrices characterized by a scalar SE equation.  The main intuition for such algorithms is that using the SVD of the measurement matrix , $\*\Phi = \*U\*S\*V^T$, the right-orthogonal random component, $\*V^T$ can be decoupled from the poorly conditioned component, $\*U\*S$ which is dealt with via a linear MMSE estimator component within the VAMP iteration \cite{rangan2016}. While this significantly increases the class of matrices for which AMP techniques can be applied it still requires the calculation of the SVD. For large imaging problems, such as 2D or 3D CT imaging such a calculation is not practical as the operators themselves are computed on the fly and not stored in matrix form. In contrast, the approach we propose here similarly removes the ill-conditioning, but by right-multiplying by an easy to compute preconditioner, thus making it more attractive to large scale CT imaging applications. Another difference from VAMP is therefore that our preconditioner modifies the signal space and thus the signal model is defined in the preconditioned space rather than the original image space.

Finally, it is useful to draw a link with the existing literature on model based iterative reconstruction (MBIR) for CT imaging. Current state-of-the-art MBIR solutions for CT are based on minimizing a regularized negative log-likelihood (NLL) cost function or its approximation using penalized weighted least squares, \cite{erdogan1999,elbakri2002,nuyts2013}, which can be interpreted as a Bayesian maximum a posteriori (MAP) estimator. This MAP framework can also be modified to incorporate denoising functions using the Plug-and-play (PnP) framework \cite{ramani2016,venkatakrishnan2013}. In contrast, our proposed algorithm takes the MMSE estimator perspective on AMP and we analyse the equations of the SE prediction associate with the MMSE formulation of GAMP. Furthermore, as MAP estimation reduces to an optimization problem,  the conditioning effects of the noise and system models are intertwined such that typical preconditioning has only a limited benefit. Using a preconditioned GAMP framework allows us to decouple these two effects.

\subsection{Notation}
Matrices or discrete operators and column vectors are written respectively in capital and normal boldface type, i.e. $\*A$ and $\*a$ to distinguish from scalars and continuous variables written in normal weight. $(\cdot^T)$ and $(\cdot^H)$ refer respectively to the transpose and the conjugate transpose of a matrix and $\*1$ refers to a vector of ones. Non-random quantities and random realizations are not distinguished typographically while random variable are written with capital letters. The conditional probability density function of $\*y$ given $\*x$ is denoted alternatively by $p_{Y|X}(\*y|\*x)$ or $p(\*y|\*x)$. A Gaussian random variable $\*x$ with vector mean $\*a$ and isotropic variance $b$ is denoted by $\*x\sim\mathcal{N}(\*a,b\*I)$. $\langle\*a,\*b\rangle=\*b^T\*a$ refers to the vectors inner product.

\subsection{Structure of the Paper}
The remainder of this paper is structured as follows: Section
\ref{sec:CTmodel} briefly describes the physical model of transmission X-ray CT  from the continuous to discrete domain, introduces the Poisson non linear noise model and the approximations that lead to the Plug-and-play statistical CT reconstruction problem. The section concludes with a discussion on the effects of the system and noise models on the conditioning of the problem. 
Section \ref{sec:revAMP} reviews the original AMP algorithm for CS reconstruction, while Section \ref{sec:DCTGAMP} presents the proposed D-GAMP-CT algorithm highlighting the innovations which consist in utilizing the preconditioning for the Radon operator and incorporating the non linear CT Poisson noise model. Furthermore, we show empirical results for the SE of D-GAMP-CT. Finally, in Sections \ref{sec:numphantom} and \ref{sec:results} comprehensive results of D-GAMP-CT on a numerical phantom and experimental acquisitions of cargo luggage are shown together with a comparison of its performance with state-of-the-art algorithm for model-based CT reconstruction. 

\newpage

\section{X-ray Computed Tomography Model}\label{sec:CTmodel}

\subsection{Continuous-to-discrete model}
X-ray CT produces images of attenuation coefficients of the object or patient being scanned. A typical geometry of a CT scanner involves an incoherent source of X-ray radiation and a detector array recording the intensity of the radiation exiting the object along a number of paths. If the intensity of the source of radiation, $I_{0}$, passing through the object is known, then Beer's law provides the expected intensity after transmission, $I_{i}$ of the $i$-th ray as: 
\begin{equation}\label{BeerLaw}
I_{i} = I_{0}e^{-\int_{L_i}\mu(\vec{\nu})dl}+\epsilon_i
\end{equation}
where $\int_{L_i}\cdot dl$ is the line integral along $L_i$ which is the path of the $i$th ray through the object from the source to the detector, $\mu(\vec{\nu})$ is the spatial distribution of attenuation and $\epsilon_i$ models the scatter and other background noise in the $i$th measurement. Equation (\ref{BeerLaw}) assumes a monoenergetic X-ray source which does not usually hold in practice. However, a common effective strategy for dealing with this consists of applying a polychromatic-to-monochromatic source correction pre-processing step \cite{whiting2006}, and in the rest of the paper we will therefore assume that we have a monoenergetic source or that it has already been appropriately corrected.

To obtain a discrete model, we should approximate the continuous attenuation function, $\mu(\vec{\nu}) \in L_2(\Real^2)$, here defined over the 2D domain, using a finite basis expansion:
 \begin{equation}\label{eq:paramdiscrete}
    \mu(\vec{\nu}) \approx \sum_{j=1}^{N} \mu_j b_j(\vec{\nu})
\end{equation}
where $\bm{\mu} = [\mu_1,\ldots, \mu_{N}]^T$ is the vector of attenuation coefficients and $b_j(\vec{\nu})$ define the $N$ basis functions associated with a discrete sampling on a $\sqrt{N} \times \sqrt{N}$ Cartesian grid. 

\noindent According to the parameterization in Eq. (\ref{eq:paramdiscrete}), the line integral becomes a summation:
\begin{equation}\label{eq:linCT}
    \int_{L_i}\mu(\vec{\nu})dl \approx \sum_{j=1}^{N} \mu_j \int_{L_i}b_j(\vec{\nu})dl = \sum_{j=1}^{N} a_{ij}\mu_j.
\end{equation}
\noindent where $a_{ij}$ represents the $i,j$ element of the system matrix describing the line integral along the $i$-path from source through object at pixel position $j$ onto each detector. Repeating this over all lines defines the full view linear tomographic system matrix $\*A = [a_{ij}]$, where we assume that a sufficient density of lines has been taken such that the operator, $\*A$, is one-to-one and hence invertible on its range, e.g. \cite{averbuch2001}. The matrix $\*A$ is constructed as an over-determined matrix of dimensions $J\times N$ where $J$ is the product between the number of detectors $N_{dec}$ and the number of projections $N_{\theta}$. 

Considering the sparse view scenario, the sub-sampled CT operator can now be represented as the application of a row sub-selection operator, $\*S$ of dimensions $M\times J$, to ${\*A}$, such that the linear part of the measurement system can be described in matrix form by 
\begin{equation}\label{eq:Phi_mtx}
{\bm\Phi} = \*S{\*A}\in\mathbb{R}^{M\times N}
\end{equation}

\noindent with an effective undersampling ratio given by $M/N$. 

In the case of normal exposure, the transmitted photon flux, $I_{i}$, follows a Poisson distribution. Using the discrete parameterization, Eqs. (\ref{eq:paramdiscrete}) and (\ref{eq:linCT}), we obtain the following discrete generalized linear model:
\begin{equation}\label{eq:Poisson_CT}
    Y_i  \sim \mathrm{Poisson}\left\{I_0 e^{-z_i}+\epsilon_i \right\}, ~ i = 1, \ldots, M
\end{equation}
where $z_i$ represents the discrete (linear) projection of the $i$th ray such that, $\*z = \*\Phi\bm{\mu}$.

\subsection{Sparse view CT reconstruction}

The sparse view CT reconstruction problem aims to estimate the attenuation coefficients, $\bm{\mu}$, from the measurements $\*y = [y_1, \ldots, y_M]^T$ subject to Eq. (\ref{eq:Poisson_CT}) and any additional regularization. The negative log-likelihood (NLL) function for (\ref{eq:Poisson_CT}) given $\*y$ has the form \cite{elbakri2002}:
\begin{equation}\label{eq:NLL}
    -L(\bm\mu) = \sum_{i=1}^{M}\Big\{y_i\log\Big[I_0 e^{-[\bm\Phi\bm{\mu}]_i}+\epsilon_i \Big] - \Big[I_0e^{-[\bm\Phi\bm{\mu}]_i} + \epsilon_i \Big]\Big\}.
\end{equation}
In the case of high/normal exposure a common practice is to use a quadratic approximation of Eq. (\ref{eq:NLL}) which leads to a Weighted Least Squares (WLS) approximation \cite{elbakri2002} based on taking the logarithm of the data, $l_i = \log\Big(\tfrac{I_0}{y_i-\epsilon_i}\Big)$. This is equivalent to observing $\*z$ corrupted with a data-dependent Gaussian noise, $\*e$, with inverse covariance $\*W = \mathrm{diag}\Big[\tfrac{(y_i-\epsilon_i)^2}{y_i}\Big] $:
\begin{equation}
\*l = \*z +\*e = \*\Phi \bm\mu +\*e
\end{equation}
The NLL can then be approximated as: 
\begin{equation}\label{eq:PWLS}
 -L(\bm{\mu}) \approx \mbox{const.} + \Big(\*\Phi \bm\mu -  \*l \Big)^T \*W \Big(\*\Phi \bm\mu -  \*l \Big).
\end{equation}
For low dosage the logarithm cannot be utilized since the argument may not be non-negative, therefore Eq. (\ref{eq:NLL}) has to be used.

\subsection{Conditioning in sparse view CT}\label{subsec:PP}

It is instructive to consider the issues in minimizing (\ref{eq:PWLS}). Most popular reconstruction algorithms solve a regularized form of (\ref{eq:PWLS}) to further incorporate prior information of the image to be reconstructed:
\begin{equation}\label{eq:PWLSm}
\min_{\bm\mu\in\mathbb{R}^N_+} \frac{1}{2}||\*y - \bm\Phi\bm\mu||_{\*W}^2 + \lambda P(\bm\mu)       
\end{equation}
with $P$ usually a convex and possibly non-smooth regularization function. Assuming (\ref{eq:PWLSm}) is convex, many first order methods, like FISTA, can be applied to solve the optimization problem. Furthermore, it is possible to integrate denoising priors, such as BM3D or deep learning-based denoisers into ADMM or other algorithms using the non-convex Plug-and-play PP-WLS framework. However the convergence rate of such methods is highly dependent on the conditioning of the problem which in turn is a function of the Lipschitz constant of the data fit term $L=\sigma_{max}(\bm\Phi^T\*W\bm\Phi)$ where $\sigma_{max}$ is the maximum eigenvalue. A large value of $L$ requires the use of a small step-size to ensure stability and results in slow convergence. 

If the weighting matrix $\*W \propto \*I$, we are faced with the challenge of finding a preconditioner for the system matrix $\bm\Phi = \*S \*A$ and fortunately there exist good preconditioners for this scenario based on the geometry of the tomographic problem.
For example, this has been used in \cite{nilchian2013}  where solutions for the direct inversion of $\*A$ through a filter back projection operator are exploited. Indeed, both $\*W$ and $\bm\Phi^T\bm\Phi$ are separately easy to precondition. 

However, together, as in the PP-WLS framework, it is much more challenging. One approach that has been proposed \cite{kim2015} is to construct a diagonal preconditioner, $\mathbf{D}$, that majorizes the matrix, $\bm\Phi^T\*W\bm\Phi$:
\begin{equation}\label{eq:D}
    \mathbf{D} = \mathrm{diag}\left(\bm\Phi^T \*W\bm\Phi\mathbf{1}\right)>\bm\Phi^T\*W\bm\Phi
\end{equation}
This solution exploits the non-negativity property of the measurement matrix $\bm\Phi$. Unfortunately, this type of preconditioner does not take into account the geometric structure in the system and therefore typically only provides modest speed improvements. 
Moreover adaptive methods for estimating the Lipschitz constant of accelerated first order solvers for composite minimization \cite{florea2017}, backtracking line search \cite{calatroni2017} or adaptive restart \cite{o2015} have been proposed. For the problem of CT reconstruction, heuristic line search techniques have been used as in \cite{jorgensen2011}. However, these methods do not fundamentally change the Lipschitz constant and so may still be limited by ill-conditioning.

We will see that the GAMP framework enables us to avoid such problems by decoupling the measurement and noise components of the system. We are therefore able to exploit a preconditioner designed specifically for $\*A$ which we detail next.

\subsection{Preconditioning of the Radon operator}\label{sec:precon}

The aim is to replace the poorly conditioned operator, $\*A$ with a new operator, $\widetilde{\*A}$, that has a small condition number, i.e. it is a nearly tight frame, by mapping to a preconditioned image space. For 2D CT with parallel projections or fan-beam with appropriate resampling, our proposed solution is to use a cone filter applied in the image domain that amplifies high spatial frequencies, as has previously been used to accelerate reconstruction convergence of Conjugate Gradient solver for PWLS \cite{fessler1999}, \cite{ramani2012}(Sec. III D). 
In order to construct a discrete preconditioner, while staying geometrically exact to the continuous setting we follow the work \cite{Clinthorne1993}; since the operator $\*A^T\*A$ is approximately block-Toeplitz for shift invariant imaging CT problem, circulant preconditioners also called "Fourier" diagonalizing preconditioners have been applied to both image restoration problems \cite{nagy1996}(Sec. III A), and also shift-variant CT problems \cite{fessler1999}.

The continuous 2D X-ray Transform is a linear operator $\+A: L_2(\mathbb{R}^2)\rightarrow L_2([0, \pi) \times\mathbb{R})$ which computes the line integral of a function in the 2D input space. The Fourier central slice Theorem states that
\begin{equation}
\+A = \+F^{-1}_{\gamma} \Omega_{\omega^{-1}}\+F
\end{equation}

\noindent where $\+F$ is the 2D Fourier transform (FT), $\Omega$ is the coordinate transform operator from Cartesian to polar coordinates, $\omega^{-1} = (\gamma\cos\delta, \gamma\sin\delta)$ and $\+F^{-1}_{\gamma}$ is the inverse 1D FT with respect to $\gamma$. The output of the linear operator is the sinogram that is a function of $\delta$ and the polar space variable $\rho$. Both $\+A$ and $\+A^T\+A$ are normal-convolutional operators since
\begin{equation}\label{eq:Acontinous}
    \+A^T\+A\mu = \mu \circledast \frac{1}{|\gamma|} 
\end{equation}

\noindent and in the frequency domain
\begin{eqnarray}\label{eq:Gram_continous}
\+A^T\+A &=& \+F^H (\Omega_{\omega^{-1}})^T(\+F^{-1}_{\gamma})^H \+F^{-1}_{\gamma} \Omega_{\omega^{-1}}\+F \stackrel{(a)}{=}\+F^H (\Omega_{\omega^{-1}})^T \Omega_{\omega^{-1}}\+F\\
&\stackrel{(b)}{=}&\+F^H |\mathrm{det} J_{\omega}|\+F = \+F^H\+D\left(\frac{1}{|\rho|}\right)\+F\nonumber
\end{eqnarray}
\noindent where (a) follows from $\+F_{\gamma}$ being an unitary operator and (b) derives from the back-projection CT filter formulation where $J_{\omega}$ defines the Jacobian of $\omega$ and $\+D\left(\frac{1}{|\rho|}\right)$ is the diagonal polar Fourier space operator. From the continuous to the discrete domain, the Fourier-based Radon transform can be written as \cite{matej2004iterative,o2006fourier}: 
\begin{equation}\label{eq:Radon_op}
\*A = \*F^{-1}_{\gamma} \bm\Omega_{\bm\omega^{-1}}\*F
\end{equation}
where $\*F$ is the 2D unitary discrete Fourier transform operator which takes as input the image $\bm\mu$ of dimensions $I\times I$, with $I = \sqrt{N}$. The operator $\bm\Omega_{\bm\omega^{-1}}$ performs a discretized version of the continuous coordinate transform in Eq. (\ref{eq:Acontinous}) which outputs a matrix of polar coordinate samples that are equally-spaced along $\rho$ at the discrete locations $\{i\Delta_{\rho}\}$ for $i=-\frac{I}{2},\ldots,\frac{I}{2}-1$. The degree of approximation between continuous to discrete domain within the Fourier-based approaches is determined by the non-uniform interpolation in the frequency space \cite{potts2000new}. In \cite{fessler2003nonuniform}, a min-max analysis provides the interpolator that minimizes the worst case error. Whilst no analytical formula exists for specifying the optimal choice of the scaling function, the Kaiser-Bessel interpolation kernel can provide good compromise between accuracy and simplicity.

The non-uniform FFT operator $\bm\Omega_{\bm\omega^{-1}}\*F$ takes as input the $I\times I$ input image matrix and output a matrix of dimensions $N_{dec}\times N_{\theta}$; $\*F_{\gamma}$ applies the 1D unitary discrete Fourier transform (DFT) matrix separately to each of the radial lines vectors of dimension $N_{dec}$ and it is defined as the Kronecker product between a 1D DFT matrix $\*F_1$ and the identity matrix, i.e. $\*F_{\gamma} = \*F_1\otimes \*I_{N_{\theta}}$. Therefore, the final output is a vector of dimensions $J=N_{dec}\cdot N_{\theta}$ where $N_{dec}$ is the number of detectors and $N_{\theta}$ is the number of angles (or number of projections) in agreement with (\ref{eq:Phi_mtx}). This formulation has the advantage of being approximately, up to the gridding interpolation, one-to-one. Since $\bm\Omega_{\bm\omega^{-1}}\*F$ and hence $\*A$ are poorly conditioned, one solution is to replace $\*A$ with a better conditioned modified transform $\widetilde{\*A}$. This is equivalent to working in a new preconditioned signal space, $\*x = \*V \bm\mu$ via real-valued the linear transform
\begin{equation}\label{eq:PPFT preconditioner}
\*V = \*F^{-1}\*C^{\frac{1}{2}} \*F 
\end{equation}
where $\*C=\mathrm{diag}\left(\frac{1}{\sqrt{a^2+b^2}}\right)$ is a diagonal matrix, on the vector space, that normalizes the FT components by the sampling rate relative to the Cartesian samples $(a,b)$, with $|\rho| = \sqrt{a^2+b^2}$, which corresponds to the point spread function of the $\*A^T\*A$ at a one-pixel point source located at the center of the field-of-view \cite{Clinthorne1993}(V.A). By applying $\*V$ as right preconditioner for $\*A$, we obtain the following expression 
\begin{equation}\label{eq:Aprecon}
\tilde{\*A}=\*A\*V^{-1} = \*F^{-1}_{\gamma} \bm\Omega_{\bm\omega^{-1}}\*F\*F^{-1}\*C^{-\frac{1}{2}} \*F  = \*F^{-1}_{\gamma} \bm\Omega_{\bm\omega^{-1}}\*C^{-\frac{1}{2}} \*F
\end{equation}
It is worth noting that the operators defined in both Eqs. (\ref{eq:Radon_op}) and (\ref{eq:Aprecon}) and the linear transform $\*V$ can be seen to remain real valued through the usual conjugate symmetry arguments. 
Since the operator $\*C$ is symmetric and $\*F_1$ is an orthogonal operator, the mapping from image to image results
\begin{eqnarray}\label{eq:GramATA}
\tilde{\*A}^T\tilde{\*A} &=& (\*A\*V^{-1})^T(\*A\*V^{-1}) = \*V^{-T}(\*A^T\*A)\*V^{-1}\nonumber\\
&=& \*V^{-T}[\*F^H(\bm\Omega_{\bm\omega^{-1}})^T(\*F^{-1}_{\gamma})^H \*F^{-1}_{\gamma} \bm\Omega_{\bm\omega^{-1}}\*F]\*V^{-1} \stackrel{(a)}{=} \*V^{-T}[\*F^H(\bm\Omega_{\bm\omega^{-1}})^T \bm\Omega_{\bm\omega^{-1}}\*F]\*V^{-1}\nonumber\\
&\stackrel{(b)}{=}& \*F^{-1}\*C^{-\frac{1}{2}}\*F\left[\*F^H \*C \*F\right]\*F^{-1}\*C^{-\frac{1}{2}}\*F = \*I
\end{eqnarray}

\noindent where (a) comes from $\*F_{\gamma}$ being 1D unitary discrete Fourier transform matrix and (b) follows from the linear transformation in Eq. (\ref{eq:PPFT preconditioner}). The operator $\tilde{\*A}^T\tilde{\*A}$ is real, symmetric and positive definite. Both the matrix preconditioner in Eq. (\ref{eq:PPFT preconditioner}) and its inverse have fast $\Order(I \log I)$ implementations. Other Fourier based preconditioners could have been chosen like the Pseudo Polar FT based left preconditioner. While it has the advantage that the operator is assured to be one-to-one and empirically the singular value spread of $\tilde{\*A}$, the left preconditioning (in the measurement space) changes the statistical noise model. 

For sparse view CT, the row sub-sampling operator $\*S\in\mathbb{R}^{M\times J}$ is applied, such that the overall linear measurement system can be expressed by 
\begin{equation}
\tilde{\bm\Phi} = \*S\tilde{\*A}\in\mathbb{R}^{M\times N}.
\end{equation}

\noindent An important consequence of applying such preconditioning is that the image prior to be used in the GAMP reconstruction framework needs to be defined on $\*x$ in the preconditioned space. It will also be necessary to apply a final post-processing step to map the estimated vector, $\*x$, back into the image domain $\bm\mu$.

\section{Review of the Generalized Approximate Message Passing algorithm}\label{sec:revAMP}

\noindent In this Section, we review the formulation of the GAMP algorithm proposed in \cite{rangan2011} which is a generalization of the original AMP algorithm \cite{donoho2009}. 
AMP belongs to a families of iterative algorithms for solving linear systems of the type in Eq. (\ref{eq:PWLS}) based on different Gaussian approximations of loopy Belief Propagation. In this respect AMP, S-AMP, VAMP represent alternative ways to perform variational inference but all of them enjoy rigorous state evolution behavior. All these algorithms share the same iterative structure of performing a MAP or MSE estimation of the vector mean and scalar variance in the image domain and in the measurement domain. 
\begin{align*}
	\mathrm{MAP\; or\; MSE} &- \mathrm{input\; domain}   & \mathrm{MAP\; or\; MSE} &- \mathrm{measurement\; domain} \\
	\*x &= g_{in}(\*r, \tau_r)         & \*s &= g_{out}(\*p, \tau_p)             \\
	\tau_x & = \tau_r g'_{in}(\*r, \tau_r)  &  \tau_s &= -g'_{out}(\*p, \tau_p)  
\end{align*}		
The difference between the algorithms relies on how the mean and variances are computed, i.e. the functions $g_{in}$, $g_{out}$ together with the vectors $\*r, \*p$ and to which classes of random measurement matrices they can be applied. We develop our framework based on the GAMP formulation which is detailed in the following and we will describe how it differs from the VAMP algorithm. 
GAMP considers a class of generalized linear Bayesian inference problems, precisely estimating an unknown high dimensional input vector $\bm\mu\in\mathbb{R}^N$ observed by a mixing random linear operator $\bm\Phi\in\mathbb{R}^{M\times N}$ followed by a component-wise and nonlinear noise measurement model. 

In detail, the Bayesian forward model consists of an unknown random vector $\bm\mu$ generated from a prior separable distribution $p(\bm\mu)=\prod_{i=1}^N p(\mu_i)$; the input vector is then multiplied by a  measurement matrix $\bm{\Phi}$ whose elements are i.i.d. random Gaussian distributed $\mathcal{N}\left(0,\frac{1}{M}\right)$, i.e. $\*z = \bm\Phi\bm\mu$. Finally each component of the vector $\*z$ generates a nonlinear output $y_j,\, j=1, \ldots, M$ described by a conditional probability distribution (or likelihood) $p_{y|z}(\*y|\*z)$. 

Given the fully connected graphical model with arbitrary separable prior and separable likelihood, GAMP is an efficient and tractable message passing method based on a Gaussian approximation of loopy belief propagation (BP) in the large system limit. GAMP is constructed as an iterative algorithm which sequentially estimates the vector mean associated with samples $\bm\mu$ and $\*z$ and the scalar second order statistic (variances).  By construction, GAMP can perform Max-Sum loopy BP for approximate MAP estimation, or Sum-Product loopy BP computing approximate MMSE estimates; we will focus on the latter estimation problem in this paper.
 
GAMP algorithm converts the vector MMSE estimation problem to a sequence, indexed by $t$, of scalar MMSE estimations in the input signal and measurement domain, based on the large system limit assumption. Algorithmically, given the linear estimate $\*z^t = \bm\Phi\bm\mu^t$, GAMP employs a MMSE estimator of $\*z^t$, which results from a Gaussian approximation of the sum-product loopy BP on the dense graph (induced by $\bm\Phi$), and it propagates these means and isotropic variances estimates backward through $\bm\Phi$ to give a noisy estimate for $\bm\mu$. Then, the algorithm performs a MMSE estimate of $\bm\mu$ and propagates it forwards onto the measurements again. In order to approximately perform sum-product loopy BP and to obtain the MMSE estimates, GAMP provides a framework to construct two scalar functions in the input and measurement domain, $g_{out}(\cdot)$ and $g_{in}(\cdot)$ respectively. 
We review how to construct the function $g_{out}(\cdot)$ in the measurement domain; we consider the conditional probability distribution 
\begin{equation}\label{eq:nonlin_Lik}
p(\*z^t|\*p^t,\*y,\tau_p^t) \propto e^{\log p_{Y|Z^t}(\*y|\*z^t)} e^{-\frac{1}{2 \tau_p^t} (\*z^t - {\*p^t})^T(\*z^t - {\*p^t})} 
\end{equation}
which can be interpreted as the posterior density function of the random variable $\Xi^t\sim\mathcal{N}(\*p^t, \tau_p^t\*I)$ with observation $Y\sim p_{Y|Z^t}(\*y|\*z^t)$ where $Z^t$ is a random variable associated with the linear estimate whose instance is $z^t$. By construction of the approximate sum-product loopy BP, the messages $\Xi^t\sim\mathcal{N}(\*p^t, \tau_p^t\*I)$ are Gaussian with scalar variance and the mean is defined as a perturbed version of the linear estimate $\*z^t$, i.e. 
\begin{equation}
    \*p^t = \*z^t -\tau_{p}^t \*s^{t-1}
\end{equation}
\noindent where the term (perturbation) $\tau_{p}^t \*s^{t-1}$ represents the Onsager term. Given $p(\*z^t|\*p^t,\*y,\tau_p^t)$, the approximate iterative BP for the MMSE problem is achieved by computing 
\begin{eqnarray}\label{eq:MMSEm}
\*z_0^t &:=& \Expect_{p(\*z^t|\*p^t,\*y,\tau_p^t)}[\*z^t|\*p^t,\*y,\tau_p^t]\\
g_{out}(\*p^t, \*y, \tau_{p}^t) = \*s^t &:=& \frac{1}{\tau_p^t} (\*z_0^t - \*p^t)
\end{eqnarray}
\noindent where $\*z_0^t$ is the MMSE estimate of $Z^t$ given $\Xi^t$. 
The variance $\tau_s^t$ is calculated as the average of the negative derivative of $g_{out}(p^t_i, y_i, \tau_{p}^t)$ respect to $p_i\,\forall i=1,\ldots, M$ as follows
\begin{eqnarray}
\tau_{s_i}^t &=& -\frac{\partial}{\partial p_i}g_{out}(p^t_i, y_i, \tau_{p}^t) 
\stackrel{(a)}{=} \frac{1}{\tau_p^t}\Bigg[1 - \frac{\mathrm{Var}(z^t_i|p^t_i,y_i,\tau_p^t)}{\tau_p^t}\Bigg]\\
\tau_{s}^t &=& \frac{1}{M}\sum_{i=1}^M \tau_{s_i}^t\nonumber
\end{eqnarray}
\noindent where the equality (a) follows from the derivation in \cite[Appendix D]{rangan2011}. The vector mean of the linear estimate $R\sim\mathcal{N}(\*r^t,\tau_r^t\*I)$ in the input domain is
\begin{equation}
 \*r^t = \*x^t + \tau_r^t\bm\Phi^T\*s^t  
\end{equation}

Finally, to obtain an approximate MMSE vector mean and scalar variance estimates in input signal domain given $\*r^t$, the function $g_{in}(\cdot)$ has to be constructed as follows
\begin{eqnarray}\label{eq:MMSE_in}
g_{in}(\*r^t) = \bm\mu^{t+1}&=& \mathbb{E}[\bm\mu|\*r^t, \tau_r^t]\quad \\
\tau_{\mu}^{t+1} &=& \frac{1}{N} \sum_{i=1}^N\mathrm{Var}(\mu_i|r^t_i, \tau_r^t) \nonumber
\end{eqnarray}

In the next Section we focus on the main modifications we have introduced the the GAMP algorithm which concern how to include the preconditioning in the linear operator, how to calculate Eq. (\ref{eq:nonlin_Lik}) for the case of non linear Poisson noise model and extend to non separable input signal models, i.e. how to calculate (\ref{eq:MMSE_in}) without the explicit knowledge of the prior distribution of the unknown input signal.

\section{D-GAMP-CT: Denoising CT with Poisson noise based AMP}\label{sec:DCTGAMP}

The proposed algorithm for CT reconstruction is built upon the GAMP framework with the following innovation: i) incorporate the preconditioner for the Radon operator, introduced in Section \ref{sec:precon}, such that the iterative algorithm is performed in the preconditioned space together with a new operator with a smaller condition number. Furthermore, the algorithm utilises the following properties: ii) exploit the GAMP formulation (\ref{sec:revAMP}) for the non linear Poisson noise model in Eq. (\ref{eq:Poisson_CT}); iii) use a generic denoiser in the non linear step to capture the data-dependent structure of complex images \cite{metzler2014}. The benefit of employing  i) and ii) relies on the property of decoupling the measurements and noise components unlike the solution in (\ref{eq:D}).

\subsection{Preconditioning of the measurement operator}

As described in Section \ref{sec:precon}, the Radon operator (\ref{eq:Radon_op}) can be preconditioned by using (\ref{eq:PPFT preconditioner}) such that the combined operator $\tilde{\*A}$ has a condition number considerably lower than $\*A$ \cite{averbuch2008first, averbuch2008}. Combining (\ref{eq:PPFT preconditioner}) and (\ref{eq:Phi_mtx}), we define the modified system matrix, $\tilde{\bm\Phi}$, as
\begin{eqnarray}\label{eq:precondAMP_op}
\tilde{\bm\Phi}   &=& \*S\*A\*V^{-1} = \bm\Phi\*V^{-1}\\
\tilde{\bm\Phi}^T &=& \*V^{-1}\*A^T\*S^T = \*V^{-1}\bm\Phi^T\nonumber
\end{eqnarray}
The computational complexity of both operators, $\tilde{\bm\Phi}$ and $\tilde{\bm\Phi}^T$ is of order $\Order(N\log N)$, since they are defined as a composition of element-wise operators with complexity $\Order(N)$ and the FFT, with complexity $\Order(N\log {N})$. In an equivalent way, the preconditioning leads to the following change of coordinates in the signal domain within each iteration $t$: 
\begin{equation}\label{eq:precondAMP}
\bm\mu^t = \*{V}^{-1}\*{x}^t\quad\rightarrow\quad
\*{x}^t  = \*{V}\bm\mu^t\nonumber
\end{equation}

\subsection{Incorporation of the Poisson Noise Model in GAMP}\label{sec:DCTGAMPs}

We consider the sparse views X-ray CT transmission model where the input vector $\bm{\mu}\in\mathbb{R}^N$ is passed through the linear Radon CT operator together with the angular subsampling operator, that is modelled as 
\begin{equation}\label{eq:non_lin_lambda}
\lambda_a = e^{-z_a}= e^{-\left[\tilde{\bm\Phi}\*x\right]_a},\quad a = 1,\ldots, M
\end{equation}
 
\noindent where the linear term is $\*z = \*S\*{A}\bm{\mu}=\bm\Phi\bm\mu =\tilde{\bm\Phi}\*x$ from Eq. (\ref{eq:precondAMP_op}) and (\ref{eq:precondAMP}). Finally, each component $\lambda_a$ randomly generates an output component $y_a$ of the vector $\mathbf{y}\in\mathbb{Z}^M$. The conditional probability distribution of the i.i.d. random variable $Y$ given the linear measurement $Z$ is an exponential-Poisson distribution \cite{nuyts2013}
\begin{equation}\label{eq:noise_mod}
p_{Y|Z}(\*y|\*z) = \prod_{a=1}^M\frac{1}{y_a!} e^{-(e^{-z_a})} {e^{-y_a z_a}}
\end{equation}

Fig. \ref{block:GAMP} shows the block diagram of the generative measurement model; $\*x$ is the preconditioned vector which is mapped through $\*V^{-1}$ to the vector representing the CT attenuation coefficients $\bm\mu$. It is the input of the Radon system model $\*A$ and subsequently to the sparse view operator $\*S$ as described in Eq. (\ref{eq:Phi_mtx}). According to the transmission CT model an exponential non-linearity is applied to the corresponding linear measurement vector $\*z$. Finally, the Poisson likelihood $p_{Y|Z}$ models the CT noise as described in Eq. (\ref{eq:Poisson_CT}), given the linear system $\*z = \tilde{\bm\Phi}\*x$. While the expression of the likelihood $p_{Y|Z}$ in Eq. (\ref{eq:noise_mod}) relates the random variables $Y$ and $Z$, and therefore it already includes the non-linearity (\ref{eq:non_lin_lambda}), in Fig. (\ref{block:GAMP}) we have highlighted the non-linear block followed by an auxiliary likelihood $p_{Y|\Lambda}$.    

\begin{figure*}[!t]
	\centering
	\begin{tikzpicture}[node distance=1cm,scale=.75]
	\node  [text width=1em, text centered] (a1)  {};
	\node [punkt,right= of  a1,label={[align=center]below:Preconditioner}] (tar)  {$\*V^{-1}$};
	\node [punkt,right= of tar,label={[align=center]below:Radon\\measurement\\matrix}] (b) {$\*A$};
	\node [punkt,right= of b,label={[align=center]below:Subsampling\\operator}] (b1) {$\*S$};
	\node [punkt_l,right= of b1,label={[align=center]below:Non linearity}] (b2) {$\bm\lambda = e^{-\*z}$};
	\node [punkt,right= of b2,label={[align=center]below:Poisson\\noise\\likelihood}] (c) {$p_{Y|\Lambda}$};
	\node [right= of c, text width=1em] (white) {};
	\path[->] (a1.west) edge node [above] {$\*x\in\mathbb{R}^N$} (tar.west);
	\path[->] (tar) edge node [above] {$\bm\mu$} (b);
	\path[->] (b) edge node [above] {} (b1);
	\path[->] (b1) edge node [above] {$\*z$} (b2);
	\path[->] (b2) edge node [above] {$\bm\lambda$} (c);
	\path[->] (c.east) edge node [above] {$\*y\in\mathbb{Z}^M$} (white.east);
	\end{tikzpicture}
	\caption{Computed Tomography estimation model with Poisson noise model and matrix preconditioner $\mathbf{V}$ in the image domain.}\label{block:GAMP}
\end{figure*}

In this section, we describe how to perform the MMSE estimation in the measurement domain for the nonlinear CT Poisson noise model. Given $p(\*z^t|\*p^t,\*y,\tau_p^t)$ as defined in Eq. (\ref{eq:nonlin_Lik}) with $p_{Y|Z}(\*y|\*z)$ in (\ref{eq:noise_mod}) and the vector $\*p^t$ detailed in line (\ref{algo:goto1}) of the Algorithm \ref{Algo:GAMP}, the approximate iterative BP for the MMSE problem is achieved by computing 
\begin{equation}\label{eq:MMSEmm}
\*s^t(\*p^t, \*y, \tau_{p}^t) = \frac{1}{\tau_p^t} (\*z_0^t - \*p^t),\quad \*z_0^t := \Expect_{p(\*z^t|\*p^t,\*y,\tau_p^t)}[\*z^t|\*p^t,\*y,\tau_p^t]
\end{equation}

\noindent To obtain $\*s^t(\*p^t, \*y, \tau_{p}^t)$, we need to evaluate the expectation $\Expect(\*z^t|\*p^t,\*y,\tau_p^t)$ respect to $p(\*z^t|\*p^t,\*y,\tau_p^t)$, where
\begin{eqnarray}
\frac{1}{M}\log p(\*y|\*z^t) &=& -\langle\*z^t,\*y\rangle - \langle e^{-\*z^t}, \*1 \rangle - \langle\log(\*y!),\*1 \rangle \\
p(\*z^t|\*p^t,\*y) &\propto& e^{-\langle\*z^t,\*y\rangle - \langle e^{-\*z^t}, \*1 \rangle - \langle\log(\*y!), \*1 \rangle - \frac{1}{2 \tau_p^t} ||\*z^t - \*p^t||_2^2},\;\*z^t \in \mathbb{R}_{\geq 0}^M \nonumber 
\end{eqnarray}

\noindent The expectation requires solving the following ratio of integrals for each element indexed with $a=1,\ldots, M$:
\begin{equation}\label{eq:post_exp} 
\Expect[z_a^t|p_a^t,y_a,\tau_p^t] = \frac{\int_{\mathbb{R}_{\geq 0}} z_a^t e^{\log p_{Y|Z^t}(y_a|z_a^t)}  e^{-\frac{1}{2 \tau_p^t} (z_a^t - p_a^t)^2} dz_a^t}
{\int_{\mathbb{R}_{\geq 0}} e^{\log p_{Y|Z^t}(y_a|z_a^t)}  e^{-\frac{1}{2 \tau_p^t} (z_a^t - p_a^t)^2} dz_a^t}
\end{equation}

\noindent Unfortunately no closed form solution appears to exist and therefore Laplace's method \cite{tierney1986} is used to approximate the posterior mean $\*z_0^t$ and $\tau_s^t$. In Appendix \ref{sec_app:Laplace}, the calculation for $\*z_0^t$ and $\mathrm{Var}[\*z^t|\*p^t]$ is detailed. 

It is worth noting that the solution obtained by BM3D-CT-GAMP, using the Poisson noise model, is different from the solution of the regularized NLL minimization problem stated in Eq. (\ref{eq:NLL}).

\subsection{Denoising: Non-Linear Input Module}\label{sec:Denoiser}

Whilst the original GAMP algorithm was developed on a factorial (sparse) signal model, the framework has been shown to be amenable to much broader classes of estimators \cite{metzler2014, berthier2020}. 
Since the GAMP algorithm approximates the estimate for $\*x$ as a Gaussian noise corrupted version of the true signal with variance $\tau_r^t$ as in Eq. (\ref{eq:rGAMP}), it is meaningful to employ, instead of a prior-based non linear scalar function, a denoiser $D_{\tau_r^t}$ which acts as a standard non-linear mapping 
\begin{equation}
D_{\tau_r^t}(\cdot): \mathbb{R}^N\rightarrow\mathbb{R}^N,\quad \*r\longmapsto D_{\tau_r^t}(\*r)
\end{equation}
\noindent that, given a noisy signal estimate 
\begin{equation}\label{eq:rGAMP}
\*r=\*x+ \sqrt{\tau_r^t}\bm\psi
\end{equation}
\noindent with $\bm\psi\sim\mathcal{N}(\*0, \*I)$, outputs an estimate of  $\*x$. We treat $D_{\tau_r^t}(\cdot)$ as a black box estimator, i.e., we do not require knowledge of its functional form \cite{metzler2014}. 

The main reason for using a generic denoiser in the non linear step is to capture the data-dependent structure of complex images, rather than a simple factorial model, obtaining a sequence of estimates eventually converging faster to the true preconditioned signal $\*x$; this provides the flexibility in using a variety of denoisers. Given the estimated signal 
\begin{equation}\label{eq:r_input}
 \*r^t = \*x^t + \tau_r^t{\*V}^{-1}\bm\Phi^T\*s^t  
\end{equation}
\noindent which is the input of the denoiser, the output vector estimate and the scalar variance are given by
\begin{eqnarray}\label{eq:D_in}
\*x^{t+1}&=& D_{\tau_r^t}(\*r^t) \\ 
\tau_x^{t+1} &=& \tau_{r}^t D'_{\tau_r^t}(\*r^t)\nonumber 
\end{eqnarray}

\noindent where $D'_{\tau_r^t}(\cdot)$ denotes the divergence of the denoiser, which is by definition the sum of the partial derivatives with respect to each element $x_i, i=1,\ldots, N$ of $\*x$ and it is a scalar, i.e. 
\begin{equation}
D'_{\tau_r^t}(\*r) = \mathrm{div}_{\*r}(D_{\tau_r^t}(\*r)) = \frac{1}{N}\sum_{i=1}^N\frac{\partial D_{\tau_r^t}(\*r)}{\partial r_i}
\end{equation}

\noindent For input vector $\*r$ belonging to a simple class of signals $\+C$, it is possible to construct a denoiser which compute the conditional mean as in the MMSE formulation of GAMP in Eq. (\ref{eq:MMSE_in}), but for a general class of signals, the denoiser $D_{\tau_r^t}(\*r)$ does not necessarily correspond to a mean estimator. By design, the denoiser $D_{\tau_r^t}(\cdot)$ is acting in the preconditioned image space which consists of a high pass filtering of the image in the original spatial domain. The main properties of the denoiser are to be monotone, which means that the risk 
\begin{equation}
R(\*x, \tau_r^t) = \frac{1}{N}\-E\|D_{\tau_r^t}(\*x + \sqrt{\tau_r^t}\bm\psi) - \*x \|^2_2
\end{equation}	
\noindent is a non-decreasing function of $\tau_r^t$, and proper, i.e. $\sup_{\*x\in\+C} R(\*x, \tau_r^t)\leq \nu\tau_r^t$, for $\nu\in(0,1)$; this implies that given an estimate of $\tau_r^t$, it results $\tau_r^{t+1}\leq\tau_r^{t}$. Therefore, even when the input of the denoiser belongs to the preconditioned space, as in Eqs. (\ref{eq:precondAMP}) - (\ref{eq:r_input})  which highlight the fact that the noise is no more uncorrelated, traditional denoisers can still be used at the cost of a decrease in the rate or reduction of the risk. Similar arguments are used in plug-and-play framework \cite{buzzard2018} where at each iteration the noise term is in general correlated to the signal. 

In our framework, because of the design of the preconditioner as a high-pass filter, it is possible to utilize better denoisers which can handle this signal mapping; one choice that we have compared in the result is by using a modification of the proximal-based TV denoiser where the $\|\cdot \|_{(\*V^T\*V)^{-1}}$ norm is used instead of the $l_2$ norm. 

In Section \ref{subsec:proxTV}, we show that using this tailored denoiser leads to an improvement in the accuracy error only at earlier iterations, before convergence, compared to $l_2$ proximal TV map. The analytic calculation of $D'_{\tau_r^t}(\cdot)$ is often not available and it is in general data-dependent, but a good approximation can be obtained through the Monte Carlo technique. 
In \cite{ramani2008} the authors showed that given a denoiser $D_{\tau_r^t}(\cdot)$ and an i.i.d. random vector $\*b\sim\mathcal{N}(0, \*I)$, the divergence can be estimated as
\begin{equation}\label{eq:MC_SURE}
D'_{\tau_r^t}(\*r)\approx \frac{1}{N}\mathbb{E}_{\*b}\left[\frac{1}{\epsilon}\*b^T\Big(D_{\tau_r^t}(\*r + \epsilon \*b) - D_{\tau_r^t}(\*r) \Big)\right],\;\epsilon\rightarrow 0
\end{equation}
\noindent where the expectation over the random variable $\*b$ is calculated using a Monte-Carlo method, i.e. generate $K$ i.i.d. $\mathcal{N}(0, \*I)$ samples vectors, estimate the divergence for each vector and then obtain the global divergence by averaging:
\begin{equation}
D'_{\tau_r^t}(\*r) = \frac{1}{NK}\sum_{j=1}^K \*b_j^T\left(\frac{D_{\tau_r^t}(\*r + \epsilon \*b_j) - D_{\tau_r^t}(\*r)}{\epsilon}\right)
\end{equation}
Given the vectorized image lying in a high dimensional space, it has been empirically observed \cite{ramani2008} that we can accurately approximate the expected value using only a single random sample, i.e. $K=1$. In all the simulations we have used the Monte Carlo method with $K = 1$.

With this method, the calculation of the Onsager term is more efficient since it requires only one more application of the denoiser. 
\noindent Moreover, it follows from Eq. (\ref{eq:D_in}) that the denoiser $D_{\tau_r^t}(\cdot)$, introduced in Section \ref{sec:Denoiser} acts on the high pass filtered image $\*x$, whose expression is in Eq. (\ref{eq:precondAMP}).

\begin{figure*}[!t]
	\centering
	\resizebox{\textwidth}{!}{
		\begin{tikzpicture}[node distance=1.5cm,scale=1]
		\node [text width=11em, text centered] (a1)  {};
		\node [text centered, left= of a1] (a2)  {};
		\node [punkt,right= of  a1] (V)  {$\*V^{-1}$};
		\node [punkt,below= of  V] (VT)  {$\*V^{-1}$};
		\node [punkt,right= of V] (A) {$\bm\Phi$};
		\node [punkt,below= of A] (AT) {$\bm\Phi^T$};
		\node [sum,right= of A] (diff) {$-$};
		\node [punkt,right= of diff] (gout) {\footnotesize{compute} $z_0$};
		\node [above= of gout, text width=0em] (y) {};
		\node [sum,below= of diff,label=right:$\tau^t_{p}$] (prod) {$\times$};
		\node [sum,left= of VT,label=above:$\tau^t_{s}$] (prod_s) {$\times$};
		\node [sum,left= of prod_s] (sum) {$+$};
		\node [punkt,left= of sum] (D) {$D_{\tau_r^t}$};
		\node [right= of gout, text width=1em] (output) {};
		\path[->] (a1.west) edge node [above,name=x] {$\*x^t$} (V.west);
		\path[->] (V) edge node [above] {$\bm\mu^t$} (A);
		\path[->] (A.east) edge node {} (diff);
		\path[->] (prod) edge node [right] {$\tau^t_p\*s^{t-1}$} (diff);
		\path[->] (diff.east) edge node [above] {$\*p^t$} (gout.west);
		\path[->] (y.south) edge node [right] {$\*y$} (gout.north);
		\path[->] (gout.east) edge node [above, name=r] {$\*s^t$} (output);
		\path[->] (AT.west) edge node {} (VT.east);
		\path[->] (VT.west) edge node {} (prod_s.east);
		\path[->] (prod_s.west) edge node {} (sum.east);
		\path[->] (sum.north)+(0,2.08) edge node {} (sum.north);
		\path[->] (sum) edge node [above] {$\*r^{t-1}$} (D.east);
		\draw [->] (r) |- (AT.east);
		\draw [-] (D.west)  -| (a2.west) ; 
		\path[-]  (a2.west) edge node {} (a1.west) ;
		\path[->] (prod.south)+(0,-0.37) edge node {} (prod.south);
		\draw [color=gray,thick] (2,0)  (2,-2);
		\path (prod.west)+(-0.5,-1.1) node (g) {};
		\path (V.west)+(-0.5,-1.0) node (gg) {};
		\path[->] (-3,0) node (V) {};
		
		\begin{pgfonlayer}{background}[node distance=1cm]
		\node [background,fit=(a1) (a2) (D) (x) (sum) (prod_s),label=above:Denoising] (denoise) {};
		\node [background,fit=(output) (gout) (g) (y),label=above:MMSE estimator - measurement domain] (MMSE) {};
		\node [background,fit=(gg) (A) (AT) (VT),fill=red!30,label=above:Change of the signal domain] (prec) {};
		\end{pgfonlayer}
		\end{tikzpicture}}
	\caption{Block diagram of the D-GAMP-CT framework highlighting the 3 steps: 1) Denoising the signal estimate; 2) Preconditioning: change of the signal domain; 3) MMSE estimator for the non linear Poisson noise model.}\label{fig:diagramDCTGAMP}
\end{figure*}

The block diagram for the mean calculation of the proposed D-GAMP-CT algorithm is shown in Fig. \ref{fig:diagramDCTGAMP}; each iteration flow can be decomposed in 3 main steps: the MMSE estimation for the Poisson noise channel of the output vector $\*p^t$, the preconditioning, which involves a change of the signal domain, and the denoising of the signal estimate. Fig. \ref{fig:diagramDCTGAMP} graphically describes the steps for updating the mean vector variables of D-GAMP-CT algorithm listed in Algorithm \ref{Algo:GAMP}; the denoising block corresponds to lines \ref{algo:goto3} and \ref{algo:goto_x}, while the application of the preconditioning matrix is needed in line \ref{algo:goto5} and \ref{algo:goto3} and finally the MMSE estimation in the measurement domain corresponds to lines \ref{algo:goto7}, \ref{algo:goto9}, \ref{algo:goto10}.

\begin{algorithm}[!h]
\caption{D-GAMP-CT: Denoising Preconditioned Approximate Message Passing}\label{Algo:GAMP}
\begin{algorithmic}[1]
\STATE{\textbf{Initialization}: set $t=0$, $\*r^0=\*0$, $\*x^0=\*0$, $\tau_x^0 = 1$}
 \FOR{$1,\ldots, T_{max}$}   
 \STATE{\text{Step 1: Estimate in the measurement domain}} 
  \STATE{$\*z^t = \bm\Phi\*{V}^{-1}{\*x}^t\label{algo:goto5}$}
  \STATE{$\tau_p^t = \frac{1}{M}\|\bm\Phi\*{V}^{-1}\|_F^2\tau_{x}^t$}
  \STATE{$\*p^t = \*z^t -\tau_{p}^t \*s^{t-1}\label{algo:goto1}\label{algo:goto7}$}
  \STATE{\text{}}
  \STATE{\text{Step 2: Poisson noise model}}
  \STATE{$\*z_0^t = \Expect_{p(\*z^t|\*p^t,\*y,\tau_p^t)}[\*z^t|\*p^t,\*y,\tau_p^t]\label{algo:goto9}$}
  \STATE{$\*s^t \,= \frac{\*z_0^t - \*p^t}{\tau_p^t}\label{algo:goto10}$}\hspace*{15em}
  \rlap{\smash{$\left.\begin{array}{@{}c@{}}\\{}\\{}\\{}\\{}\\{}\end{array}\right\}
  		\begin{tabular}{l}MMSE estimation in the\\measurement domain.\end{tabular}$}}
  \STATE{$\tau_s^t = \frac{1}{M\tau_p^t}{{\sum_{i=1}^M}}\bigg[1 - \frac{\mathrm{Var}(z^t_i|p^t_i,y_i,\tau_p^t)}{\tau_p^t}\bigg]$}
  \STATE{\text{}}
\STATE{\text{Step 3: Estimate in the signal domain}}
\STATE{$\frac{1}{\tau_r^t} = \frac{1}{N}\|\bm\Phi\*{V}^{-1}\|_F^2\tau_s^t$}
\STATE{$\*r^t \,= {\*x}^t + \tau_r^t{\*V}^{-1}\bm\Phi^T\*s^t\label{algo:goto3}$}
  \STATE{\text{}}
\STATE{\text{Step 4: Denoising step}}
\STATE{$\*x^{t+1} = D_{\tau_r^t}(\*r^t)\label{algo:goto_x}$}\hspace*{13.3em}
\rlap{\smash{$\left.\begin{array}{@{}c@{}}\\{}\\{}\end{array}\right\}
		\begin{tabular}{l}MAP/MSE estimation in the\\ input domain.\end{tabular}$}}
\STATE{$\tau_x^{t+1} \,= \tau_r^t D'_{\tau_r^t}(\*r^t)$}
\ENDFOR
\RETURN $\bm\mu^t=\*{V}^{-1}\*x^t$
\end{algorithmic}
\end{algorithm}

\subsection{State evolution of D-GAMP-CT}\label{sec:DGAMPse}

A significant characteristic of GAMP is that the MSE performance can be precisely predicted by a scalar SE analysis, with i.i.d. Gaussian random system matrices in the large system limit \cite{rangan2010}; in particular, the GAMP SE formulation extends the AMP SE to arbitrary noise distributions. 

In addition, if a generic denoiser is used within the AMP (D-AMP) iterations as in Eq. (\ref{eq:D_in}), it is shown heuristically in \cite{metzler2014} that the MSE can be similarly predicted by the SE and, recently, a rigorous derivation of the SE for D-AMP is derived in \cite{berthier2020}. 

The heuristic SE equations for the proposed D-GAMP-CT are based on the GAMP SE derivation \cite{rangan2010} where the signal to estimate lies in the preconditioned domain and a denoiser is utilized as the non-linear input function. We should stress that a rigorous analysis for denoising GAMP has not yet been derived and that the SE analysis of D-AMP cannot be directly applied to denoising GAMP. 

\newpage 

The D-GAMP-CT SE equations follow the GAMP SE formulation \cite{rangan2010}; for the input and output vectors, we define 2 sets of vectors 
\begin{eqnarray}
\theta^t_{r} &=& (\*x, \*r^t, \*x^t)\\
\theta^t_{p} &=& (\*z,{\*z}^t, \*y,\*p^t)
\end{eqnarray}
$\theta^t_{p}$ contains the components of the true and unknown vector $\*z$, its D-GAMP-CT estimates ${\*z}^t$ and $\*p^t$ (line \ref{algo:goto1}) and the observed measurement vector $\*y$, while $\theta^t_{p}$ contains the unknown input vector $\*x$ and the D-GAMP-CT estimates $\*x^t$, and $\*r^t$ (line \ref{algo:goto3}). 
The main results in \cite{rangan2010} state that for a fixed iteration $t$ and $N\rightarrow+\infty$ the joint empirical distribution of the elements for the vectors $\theta^t_{r}$ and $\theta^t_{p}$ converges empirically with second-order moments $PL(2)$ to the random vectors as
\begin{eqnarray}\label{eq:SEasym}
\lim_{N\rightarrow +\infty} \theta^t_{r}  \overset{PL(2)}{=} \;\, \hat{\theta}^t_{r}(\xi_r^t) & = & (X, \hat{R}^t, \hat{X}^{t+1})\\
\lim_{N\rightarrow +\infty} \theta^t_{p}  \overset{PL(2)}{=}  \hat{\theta}^t_{p}(\*K^t_p) & = & (Z, \hat{Z}^t, Y, P^t)\\
\lim_{N\rightarrow +\infty}\tau_r^t =  \hat{\tau}_r^t, & & \quad \lim_{N\rightarrow +\infty}\tau_p^t = \hat{\tau}_p^t,
\end{eqnarray}
with 
\begin{equation}
	\hat{R}^t = X+V^t, \; V^t\sim\mathcal{N}(0, \xi_r^t)\quad\textrm{and}\quad \hat{X}^t=D_{\hat{\tau}_r}(\hat{R}^t)
\end{equation}
for the input vector estimation and 
\begin{equation}
(Z, P^t)\sim\mathcal{N}(0,\*K_p^t),\;\hat{Z}^t=\Expect_{p(\*z^t|\*p^t,\*y,\hat{\tau}_p^t)}[P^t, Y, \hat{\tau}_p^t]
\end{equation}
for the output vector. The SE equations in \cite[Algorithm~3]{rangan2010} produce a recursive scheme for calculating the parameters $\xi_r^t,\,\hat{\tau}_r$ of the distributions $\hat{\theta}^t_{r}$ and $\*K_p^t,\,\hat{\tau}_p^t$ for $\hat{\theta}^t_{p}$. 
In the case $p_{Y|Z}$ matches the true distribution in Eq. (\ref{eq:noise_mod}), then it results that 
\begin{equation}\label{eq:SE_matches}
\hat{\tau}_r^t = \xi_r^t = -\mathbb{E}^{-1}\left[\frac{\partial}{\partial p^t}g_{\mathrm{out}}(P^t, Y, \hat{\tau}_{p}^t)\right]
\end{equation}
where the expectation is taken over $\hat{\theta}^t_{p}(\*K^t_p)$ with 
\begin{equation}
\*K^t_p = \left[\begin{array}{cc}
\hat{\tau}_x^0                  & \hat{\tau}_x^0-\hat{\tau}_p^t\\
\hat{\tau}_x^0-\hat{\tau}_p^t   & \hat{\tau}_x^0 -\hat{\tau}_p^t
\end{array}{}\right]
\end{equation}
and $\hat{\tau}_x^0$ is set with an initial value and $\hat{\tau}_p^{t} = \beta\hat{\tau}_x^t,\quad \*K_p^t=\beta\*K_x^t$ with $\beta=\frac{M}{N}$. 
Following the derivation in \cite{Fletcher2018,rangan2010}, the error and sensitivity functions are defined. 

The error functions characterize the MSEs of the denoiser under Gaussian noise while the sensitivity functions describe the expected divergence of the estimator. The parameter $\hat{\tau}_x^t$ depends on both the error and sensitivity functions as follow. For the class of denoising functions $D_{\hat{\tau}_r}(\cdot)$ that are uniformly Lipshitz and convergent under Gaussian noise, which includes several non-separable denoisers \cite{Fletcher2018}, the sensitivity function is defined as
\begin{equation}
\hat{\tau}_x^t = \+A_{in}(\hat{\tau}_r^t, \xi_r^t) = \lim_{N\rightarrow\infty}\langle\nabla D_{\hat{\tau}_r}(\*x + \*v^t)\rangle,\quad  \*v^t\sim\mathcal{N}(0, \xi_r^t \*I)
\end{equation}
and the error function is 
\begin{equation}
\+E_{in}(\hat{\tau}_r^t, \xi_r^t) =  \lim_{N\rightarrow\infty}\frac{1}{N}\|D_{\hat{\tau}_r}(\*x + \*v^t) - \*x   \|^2,\quad  \*v^t\sim\mathcal{N}(0, \xi_r^t \*I)
\end{equation}
 
\noindent Unfortunately, the SE prediction is only valid in the random large system limit and therefore one may wonder what its relevance is in the considered CT problem. Here we argue that the empirical accuracy of the SE predictions provides an insight into the validity of D-GAMP-CT approximations when applied to such general linear models. 

In particular, we claim that the small discrepancy is mainly due to the fast that the SE is derived under the "matched" condition while we calculate the posterior mean $\*z_0^t$ and $\tau_s^t$ by the Laplace approximation in Eq. (\ref{eq:post_exp}). 

In Section \ref{sec:results}, we present empirical evidence that the SE for D-GAMP-CT based on a real CT dataset provides an excellent prediction of the actual MSE achieved by D-GAMP-CT at each iteration.

\newpage

\section{Comparison with Other Methodologies}

\subsection{Vector AMP (VAMP)}\label{subsec:VAMP}
Recently a new message passing algorithm VAMP (or its generalization VGAMP \cite{schniter2016c}) has been proposed which enjoys convergence guarantees for a larger class of random system matrices, i.e. right-orthogonal invariant. VAMP has been succesfully used in imaging application, like CT \cite{sarkar2019}, and inverse scattering \cite{sharma2019}. One difficulty within VAMP algorithm relies on the fact that its implementation requires either to compute the SVD of the system operator, or computing the covariance of the LMMSE estimator, i.e. inverting an high-dimensional symmetric matrix. Therefore, VAMP is particularly appealing for problems where it can be possible to compute the SVD of the system operator, either because it is available in matrix form, or because it can be decomposed by a fast orthogonal operator, like FFT, which leads to a fast computation of the trace of the inverse LMMSE covariance matrix. While in MRI it is possible to exploit the FFT form of the operator, unfortunately in CT, it is not generally possible to have a matrix form operator and therefore it becomes time consuming either computing the SVD off-line or estimating the inverse of the LMMSE covariance matrix. 

\subsection{VAMP with Signal Whitening}
As described in the Introduction \ref{subsec:rel_works}, different practical approaches have been proposed to handle non-random matrices within AMP or VAMP framework. An approach is the randomization of the input signal or scrambling its sample locations (or flipping its sample signs), then applying the sensing matrix on the randomized samples and finally, sub-sampling the resulting transform coefficients. Randomization methods de-correlate the signal with the sensing matrix, but generally this can be applied only with orthogonal sensing matrices, like Fourier matrix, but it does not apply to the Radon matrix for the reasons explained in Section \ref{sec:precon}. Furthermore, pre-randomizing the input might be physically not possible to implement and computationally inefficient. 
	
A more efficient method presented in \cite{Schniter2017} is based on whitening the input signal and it is designed specifically for low-frequency Fourier matrix by using an appropriate wavelet transform as whitening operator. Instead, the aim of the proposed method based on preconditioning is to construct a new operator, close to be orthogonal, in an efficient manner, i.e. by exploiting the FFT. An interesting alternative to explore in the future is to use a wavelet transform $\*T$ as whitening operator in conjunction with a Radon operator $\*A$. Several wavelets, e.g., Haar, Daubechies, can achieve $O(N)$ complexity and the overall complexity of the composition $\*A\*T$ would be $\+O(N\log N)$. Similarly, the proposed Fourier-based preconditioning enjoys low $\+O(N\log N)$ computational cost.

\subsection{Plug-and-Play Optimization}
Recent works \cite{ono2017primal, kamilov2017plug} have explored the Plug-and-Play (PnP) incorporation of modern denoisers within the explicit regularization objective function in Eq. (\ref{eq:PWLSm}), as described in Section \ref{subsec:PP}, or an alternative approach called regularization by denoising \cite{romano2017little}. PnP approaches achieve state of the art recovery results in imaging applications even if they do not minimize an explicit MAP objective function \cite{kamilov2017plug} as for the regularization by denoising approach it is not completely understood what underlying objective function is being minimised by these algorithms - see the clarifications and new interpretations presented in \cite{reehorst2018regularization}. We have compared our proposed framework with the Plug-and-Play ADMM (PnP-ADMM) optimization algorithm \cite{Sreehari2016} in Sections \ref{sec:numphantom} and \ref{sec:results}.

\section{Simulation Results with Numerical Phantom}\label{sec:numphantom}

\begin{figure*}[!h]
	\centering
	\subfloat{\includegraphics[width=.49\textwidth]{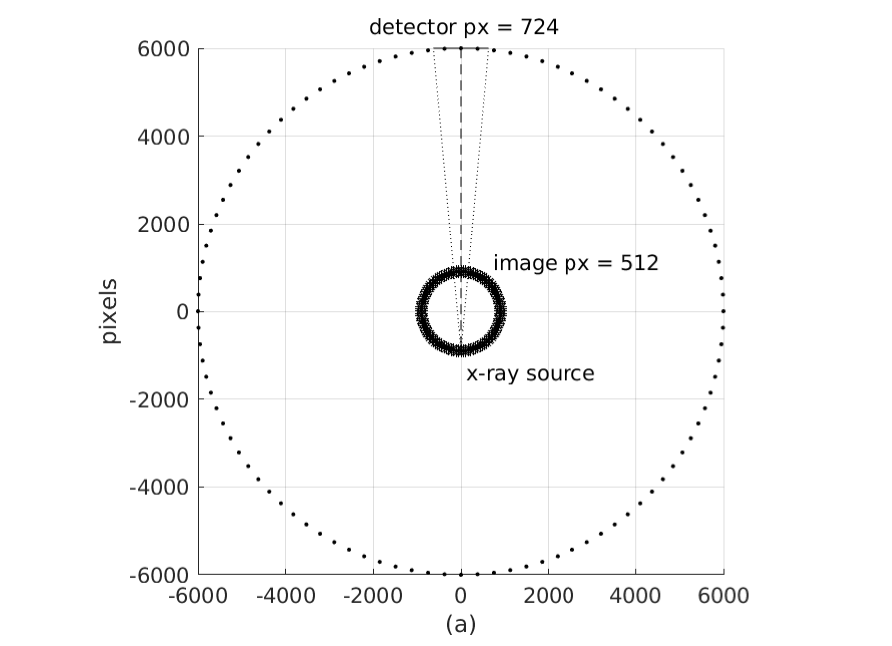}}
	\hspace{.0em}
	\subfloat{\centering\includegraphics[width=.5\textwidth]{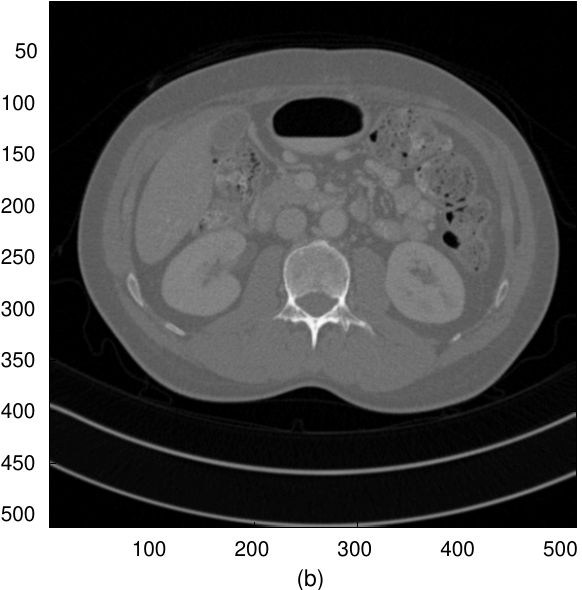}}
	\hspace{.0em}
	\subfloat{\includegraphics[width=.49\textwidth]{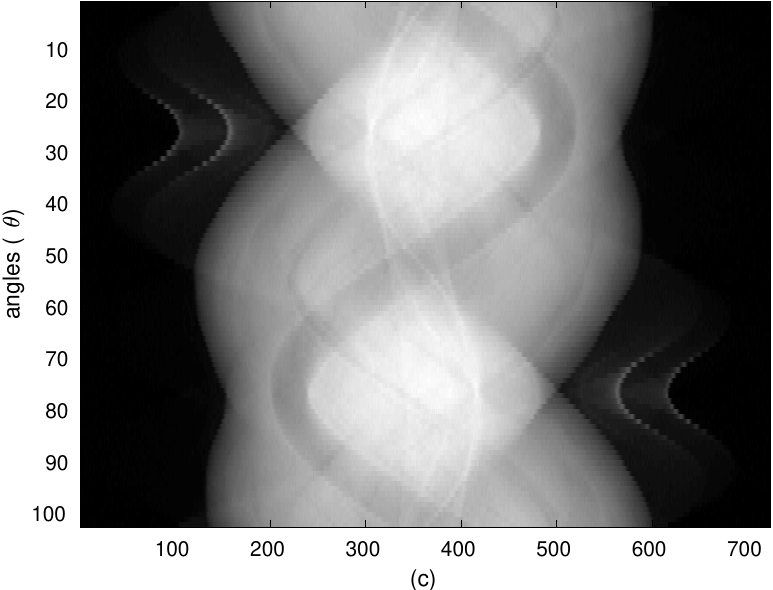}}
	\hspace{.0em}
	\subfloat{\includegraphics[width=.49\textwidth]{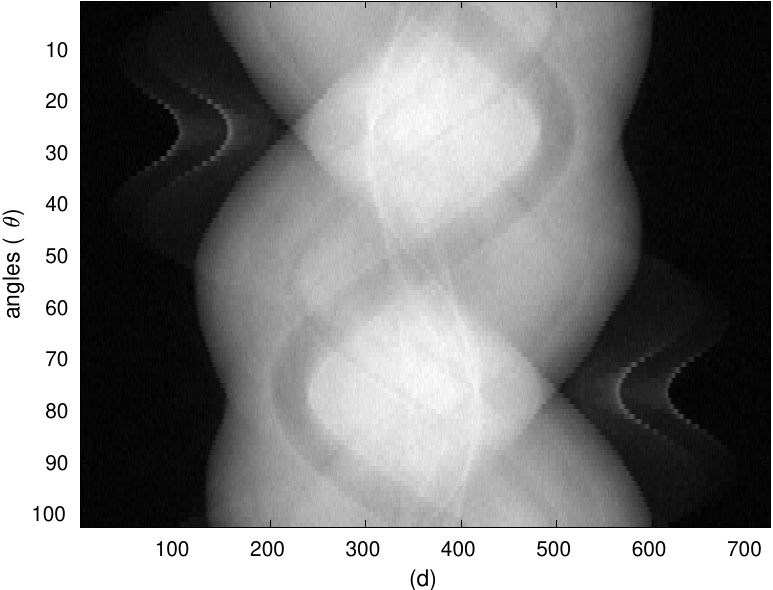}}\\
	\caption{(a) Fan-beam geometry, (b) AAPM  phantom - dataset L067\_full\_1mm, slice 170, (c) Sinogram for normal dose, $I_0 = 10^5$, (d) Sinogram for low dose, $I_0 = 10^4$.}\label{fig:system}
 \end{figure*}

We discuss the numerical results for 2D CT reconstruction using the “2016 NIH-AAPM-Mayo Clinic Low Dose CT Grand Challenge" full dataset as ground-truth; the slice 170 of dimension $(512\times 512)$ from the "L067\_full\_1mm" acquisition is shown in fig. \ref{fig:system}(b) and we have simulated a fan beam geometry, depicted in Fig. \ref{fig:system}(a); we consider $\lfloor\frac{N}{5}\rfloor = 102$ views in the sinogram domain, obtained from a regular angular undersampling of the full projection measurements (1024 views $= 2N$), resulting in, approximately, 10 times undersampling ratio. 
The CT projection and back-projection operators are implemented using the ASTRA Toolbox \cite{van2016}.

The simulations include the Poisson noise model with different levels of intensity: an initial intensity of $I_0 = 10^5$, which is referred to as normal dose in the toolbox, and $I_0 = 10^4$ for the low dose case. The sparse views sinograms, for the 2 levels of intensity, are shown in Figs. \ref{fig:system}(c)-(d) where it is worth noticing the low values in case of low dose; we will show that the Gaussian approximation of the CT noise is less effective with low beam intensity. 

\begin{figure*}[!h]
	\centering
	\subfloat{\includegraphics[width=.49\textwidth]{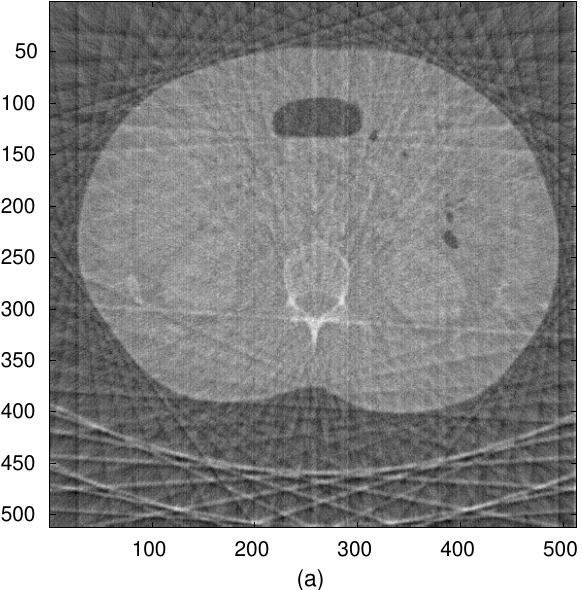}}
	\hspace{.0em}
	\subfloat{\includegraphics[width=.49\textwidth]{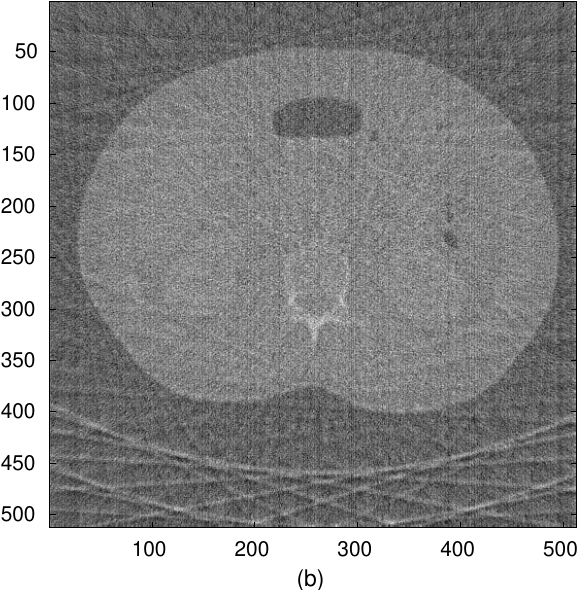}}
	\hspace{.0em}
	\subfloat{\centering\includegraphics[width=.49\textwidth]{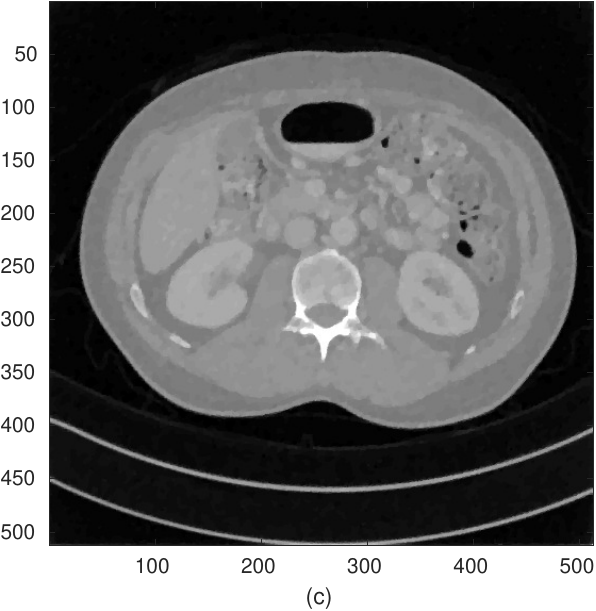}}
	\hspace{.0em}
	\subfloat{\includegraphics[width=.49\textwidth]{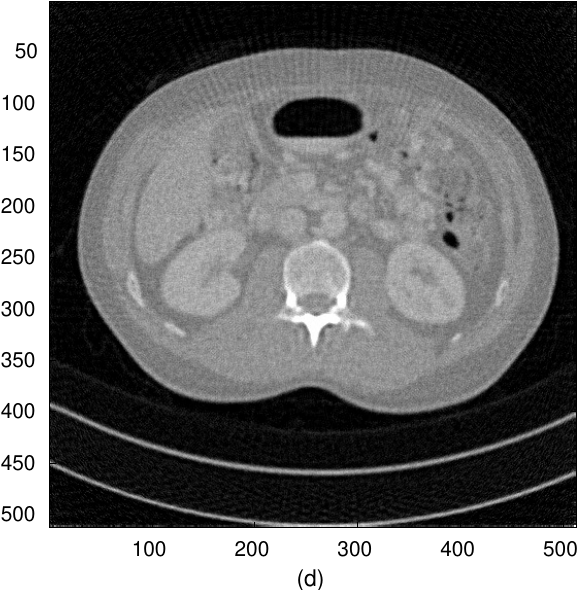}} \\
	\caption{(a) FBP Normal dose, (b) FBP low dose. Normal dose: (c) BM3D-GAMP-CT, (d) BM3D-ADMM-WLS (PnP).}\label{fig:FBPnumerical}
\end{figure*}

\subsection{Comparison D-GAMP-CT and Plug-and-play algorithms}

\begin{figure}[!h]
    \centering
    \subfloat{\includegraphics[width=.49\textwidth]{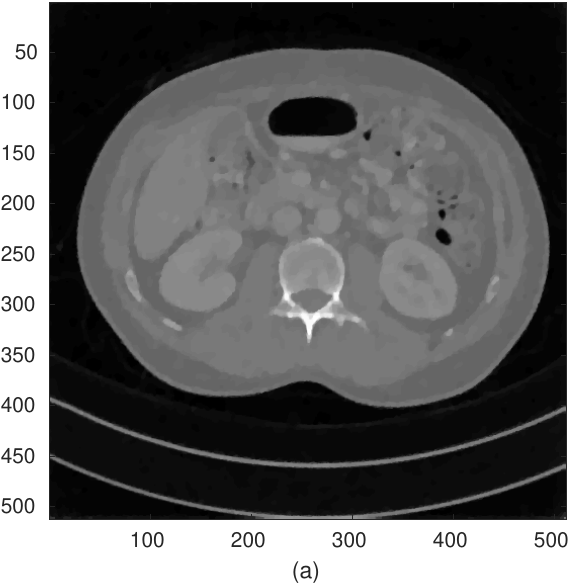}}
    \hspace{.0em}
    \subfloat{\includegraphics[width=.49\textwidth]{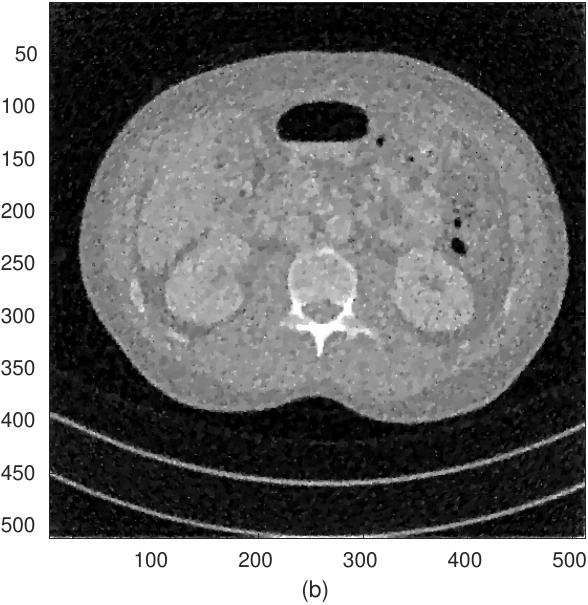}}
	\caption{Low dose: (a) BM3D-GAMP-CT, (b) BM3D-ADMM-WLS (PnP).}\label{fig:lowAMP}
\end{figure}

Figs. \ref{fig:FBPnumerical}(a)-(b) show the FBP with ramp filter, for the normal and low photon intensities, which produces very poor reconstructions with strong streaking artifacts. 
In Figs. \ref{fig:FBPnumerical}(c)-(d) are shown the reconstruction results for the normal dosage obtained using respectively BM3D-GAMP-CT algorithm \ref{Algo:GAMP}, which reaches the convergence in 10 iterations, and the BM3D plug-and-play algorithm with WLS data fidelity term, implemented using the ADMM solver (BM3D-ADMM-WLS). The BM3D-GAMP-CT is built upon the GAMP Toolbox \cite{RanganMatlab} and D-AMP Toolbox \cite{MetzlerMatlab}, while the BM3D denoiser is implemented using the Matlab toolbox \cite{danielyan2012} and the BM3D plug-and-play algorithm is implemented using \cite{venkatakrishnan2013} and is used as the reference reconstruction algorithm to compare with our proposed method. The image denoising algorithm BM3D is used as the denoiser in D-GAMP-CT since it provides good reconstruction performance and keeps computation time reasonable. It is worth noting that from Figs. \ref{fig:FBPnumerical}(c)-(d), BM3D-GAMP-CT achieves a better qualitative reconstruction compared to BM3D-ADMM-WLS (PnP), whose output retains streaking noise artifacts in the inner region probably due to the the rays intercepting the hard tissue or bones. 
In Figs. \ref{fig:lowAMP}(a)-(b) the results with low dose are shown for, respectively, BM3D-GAMP-CT and BM3D-ADMM-WLS. It is important to highlight that, in this case, the weighted Gaussian noise approximation, is not accurate due to the presence of zero values in the sinogram related in particular to the rays intercepting the bones. Taking the logarithm of the measurement leads to errors, especially in the region surrounded by hard tissue/bones; this is also confirmed quantitatively in Table \ref{tab:Phantomconf}. For a quantitative comparison, we have chosen the PSNR as the metric, defined as the ratio between the ground truth and the mean square error of the estimation, and the SSIM metric \cite{wang2004image} which scores images in the interval $[0, 1]$, where a higher index represents better quality.

Finally, it is worth comparing the proposed framework with a first order optimization solver for the regularized Poisson likelihood CT model. From the definition of the NLL function for a mixed Poisson Gaussian CT noise model in Eq. (\ref{eq:NLL}), we consider an approximation of Eq. (\ref{eq:NLL}) where we model the Poisson noise, neglecting the Gaussian noise ($\bm\epsilon\approx\*0$); therefore, the associated NLL function can be rewritten in vector form as 
\begin{equation}
L(\bm\mu) = <I_0e^{-\bm\Phi\bm{\mu}},\*1> - <\*y, \log (I_0 e^{-\bm\Phi\bm{\mu}})>  
\end{equation}

\begin{figure}[!h]
	\centering
	\subfloat{\includegraphics[width=.49\textwidth]{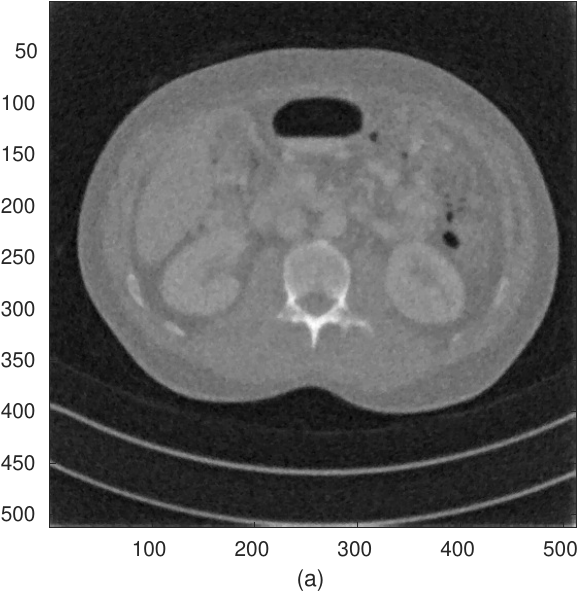}}
	\hspace{.0em}
	\subfloat{\includegraphics[width=.49\textwidth]{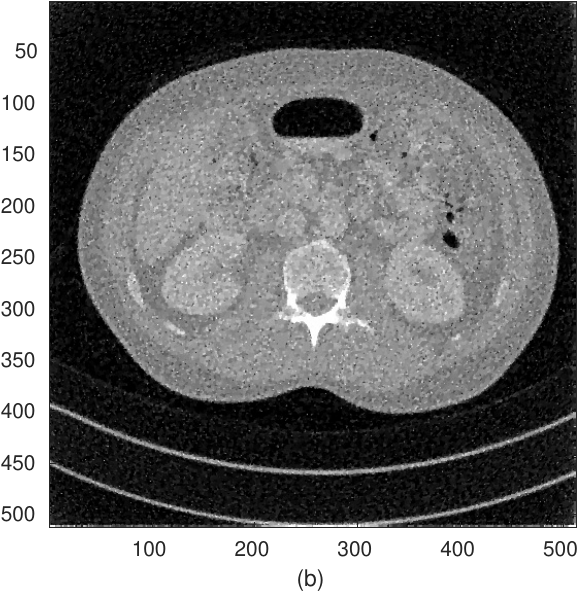}}
	\caption{BM3D-ADMM-NLL (PnP): (a) Normal Dose, (b) Low Dose.}\label{fig:ADMM_NLL}
\end{figure}

\noindent The original regularized MAP Poisson objective function can be expressed as
\begin{equation}\label{eq:costPois}
\hat{\bm\mu}_{NLL} = \arg\min_{\bm\mu}L(\bm\mu) + R(\bm\mu) + \chi_B(\bm\mu)	
\end{equation} 
\noindent where the non-negativity constraint on $\bm\mu$ is enforced by the characteristic function $\chi_B(\bm\mu)$ on the set $B=\{\bm\mu:\mu_i\geq 0, \forall i\}$. We applied ADMM with splitting to solve (\ref{eq:costPois}) as it has been proposed in \cite{ding2018}. Figures \ref{fig:ADMM_NLL}(a)-(b) show respectively the results with normal and low dose by applying BM3D-ADMM-NLL (with BM3D denoiser, the plug-and-play ADMM minimizes a different MAP cost function compared to (\ref{eq:costPois}) with no explicit regularization $R(\mu)$).

\begin{figure}[!h]
	\centering
	\subfloat{\includegraphics[width=.49\textwidth]{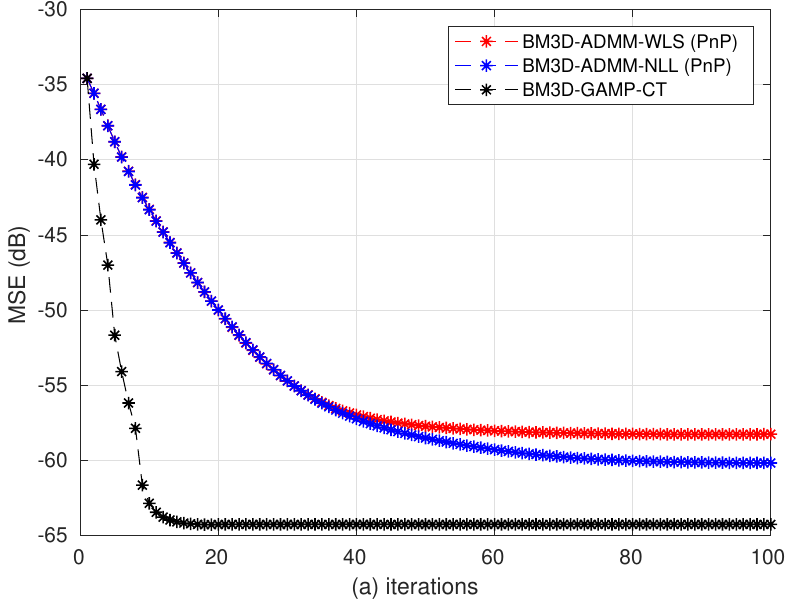}}
	\hspace{.0em}
	\subfloat{\includegraphics[width=.49\textwidth]{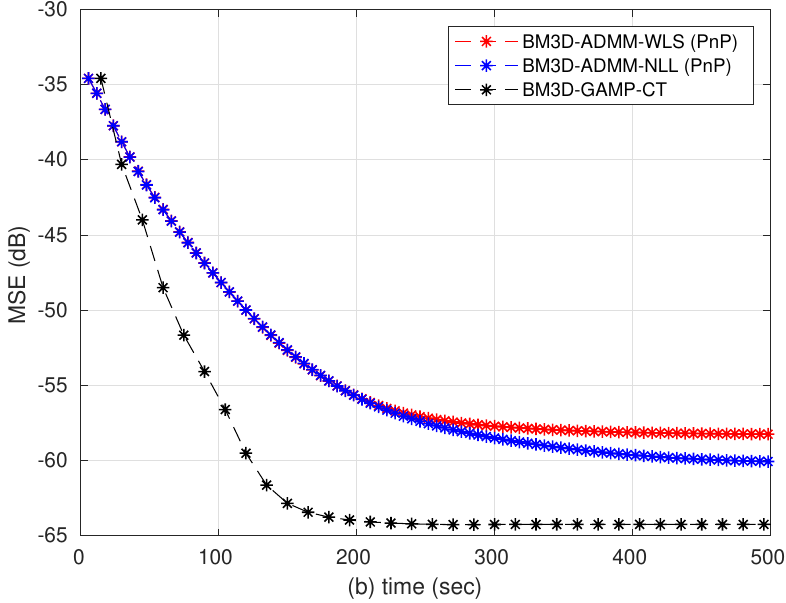}}
	\caption{Comparison between BM3D-GAMP-CT, BM3D-ADMM-NLL and BM3D-ADMM-WLS: (a) MSE vs iterations; (b) MSE vs running time (sec).}\label{fig:MSE_plot_GAMP_PP}
\end{figure}

Table \ref{tab:Phantomconf} also reports the quantitative results in terms of PSNR and SSIM for both low and high dose intensity values. The convergence plots in terms of MSE (dB) against iterations of the BM3D-GAMP-CT, BM3D-ADMM-WLS, BM3D-ADMM-NLL algorithms are shown in Fig. \ref{fig:MSE_plot_GAMP_PP}. In both normal and low dose scenario, it can be seen that BM3D-GAMP-CT produces a better quantitative reconstruction in terms of PSNR (dB) compared to BM3D-ADMM-WLS, and it requires lower total running time. For computational time evaluation, the simulations are run on an Intel\textregistered Xeon 2GHz machine using 8 cores.

\begin{table}[!h]
	\centering%
	\caption{PSNR and time comparison}\label{tab:Phantomconf}
	\vskip 0.25cm
	\begin{tabular}{lcccc}
	\textbf{Algorithms}				&	\textbf{PSNR} [dB]	 & \textbf{SSIM} & \textbf{Time}\\
	\hline
	\hline
	\textbf{Low photon intensity: $I_0=10^4$} & & \\
	\hline
	FBP						&	31.5	&  0.36 & 45 sec		\\
     BM3D-ADMM-WLS (PnP)				&	58.2	 & 0.82   & 7.8 min\\
	 BM3D-ADMM-NLL (PnP)                & 60.2 & 0.86 & 8 min\\
	 BM3D-GAMP-CT					&	64.4 & 0.95   & 4.5 min\\
	\hline
	\hline
		\textbf{High photon intensity:  $I_0=10^5$} & & \\
	\hline
	FBP						&	40.2	& 0.60  &  45 sec		\\
	BM3D-ADMM-WLS (PnP)				&	65.6	 & 0.88   & 7.8 min\\
    BM3D-ADMM-NLL (PnP)             & 67.1  & 0.91 & 8 min\\
	BM3D-GAMP-CT					& 70.4	 & 0.97   &  4.5 min\\
	\hline
	\hline
	\end{tabular}
\end{table}

\subsection{Comparison computational cost and running time}

From Algorithm \ref{Algo:GAMP} it is possible to estimate the order of complexity of the BM3D-GAMP-CT algorithm. At each iteration, 2 matrix vector multiplications are required in lines 5 and 14. For both operators $\tilde{\bm\Phi}$ and $\tilde{\bm\Phi}^T$ defined in Eq. (\ref{eq:precondAMP_op}), the matrix multiplication has a computational complexity of $\mathcal{O}(N\log N)$, since it is defined as a composition of element-wise operator of complexity $\mathcal{O}(N)$ and the operator $\*A$ of complexity $\mathcal{O}(N\log N)$. The computation of the output vector mean $\*z_0^t$ (line 9) and the variance $\tau_s^t$ does not involve matrix vector multiplications but summation of order $N$ as described in Appendix \ref{sec_app:Laplace}. 
In addition, each iteration requires the application of one denoising function in (line \ref{algo:goto_x}) whose complexity depends on the actual implementation, since it is treated as a black box and the update of the variance $\tau_x^t$ requires to compute the divergence of the denoiser $D'_{\tau_{\*r}}(\*r)$ which is implemented by Monte Carlo SURE \cite{ramani2008} whose complexity is equivalent to one denoising function. 
On the other hand, BM3D-ADMM-WLS algorithm also requires at each iteration 2 matrix vector multiplications, solving one optimization sub-problem with Conjugate Gradient (CG) and applying one denoiser (instead of complexity of 2 denoising functions required by BM3D-GAMP-CT). 
In our simulations shown in Fig. \ref{fig:MSE_plot_GAMP_PP}(b), the running time per iteration of BM3D-GAMP-CT is higher than the one of  BM3D-ADMM-WLS because even the optimized implementation of BM3D \cite{danielyan2012} has an higher cost compared to the CG solver. However, the total running time of  BM3D-GAMP-CT is lower (around 3 min as shown in Table \ref{tab:Phantomconf} and Fig. \ref{fig:MSE_plot_GAMP_PP}(b)) because fewer iterations are needed to converge. Finally, compared to BM3D-ADMM-WLS the computation for the BM3D-ADMM-NLL solver with splitting requires to additionally inverting a circulant matrix by conjugate gradient whose complexity is at least of order $\mathcal{O}(N\log N)$.

\subsection{Proximal-based TV denoiser}\label{subsec:proxTV}

\begin{figure}[!h]
	\centering
	\subfloat{\includegraphics[width=.49\textwidth]{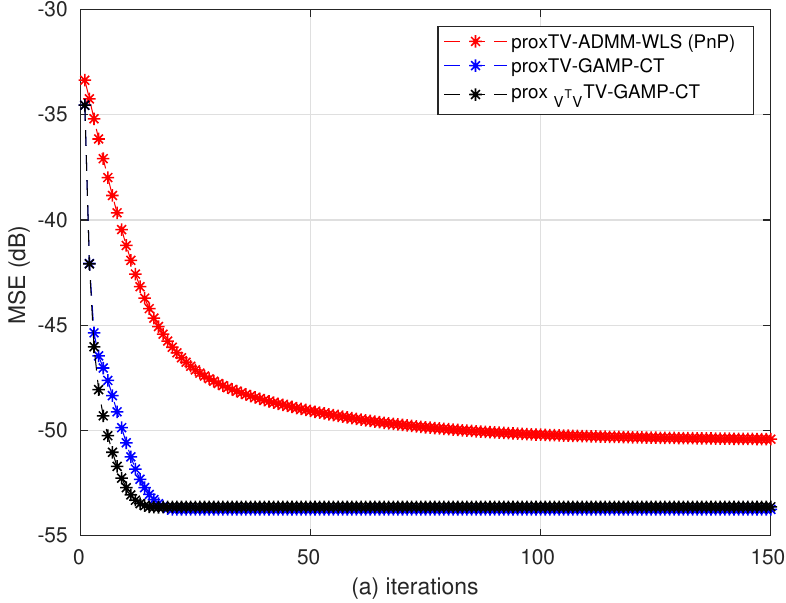}}
	\hspace{.0em}
	\subfloat{\includegraphics[width=.49\textwidth]{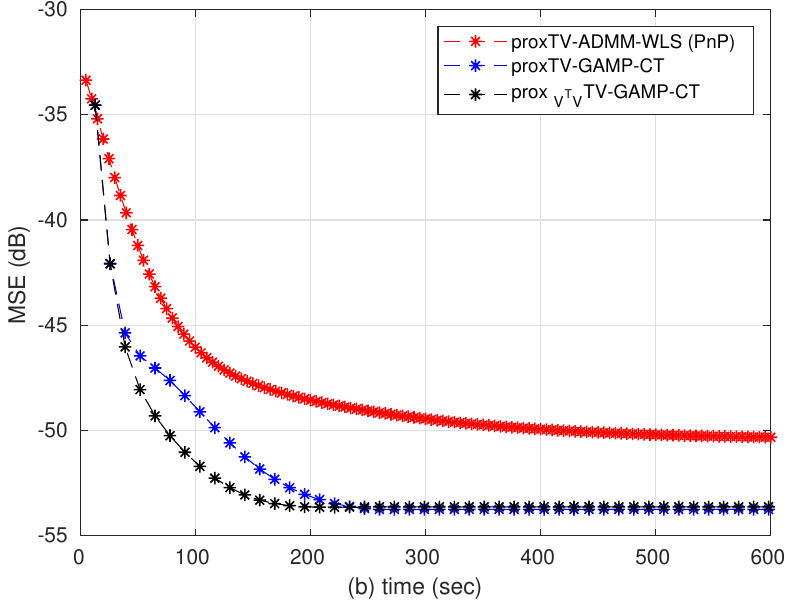}}
	\caption{Comparison between prox$_{(\*V^T\*V)}$TV-GAMP-CT, proxTV-GAMP-CT and proxTV-ADMM-WLS: (a) MSE vs iterations; (b) MSE vs running time (sec).}\label{fig:MSE_plot_TV_GAMP_PP}
\end{figure}

We show the reconstruction results using a different denoiser, the proximal Total Variation (TV) and we compare the result using a $\|\cdot \|_{(\*V^T\*V)^{-1}}$ matrix norm instead of the $L_2$ norm for the proximal-based TV denoiser to properly handle the input signal in the preconditioned domain \cite{buzzard2018}. 
Fig. \ref{fig:MSE_plot_TV_GAMP_PP}(b) shows the running time of prox$_{(\*V^T\*V)}$TV-GAMP-CT, proxTV-GAMP-CT and proxTV-ADMM-WLS; we observe that the total running time of the proxTV-GAMP-CT based algorithm is lower of the proxTV-PnP Plug-and-play and, together with the previous results with BM3D, D-GAMP-CT leads to a reduction in both number of iterations and total running time compared to the PnP approach, irrespective to the type of denoiser used. Furthermore, by comparing with the plot in Fig. \ref{fig:MSE_plot_GAMP_PP}(b) while the MSE accuracy at convergence achieved with proxTV is about 9dB worse compared to BM3D, the iteration time with proxTV is lower since the complexity of proxTV is only of order of $\mathcal{O}(N)$. 

\newpage

Interestingly, the prox$_{(\*V^T\*V)}$TV-GAMP-CT, compared to the proxTV-GAMP-CT, leads to an improvement in the accuracy error only at earlier iterations, before convergence, while both converges close to the same MSE value. This is intuitively expected because the prox$_{(\*V^T\*V)}$TV tends to better reduce the noise variance at earlier iterations since it takes into account the appropriate norm for the preconditioned space but at convergence the achieved accuracy is almost equivalent to the $l_2$ norm-based proxTV.

\subsection{Comparison between BM3D-GAMP-CT and BM3D-VAMP-CT}

\begin{figure}[!h]
	\begin{minipage}[!h]{\linewidth}
		\centering
		\includegraphics[width=.55\textwidth]{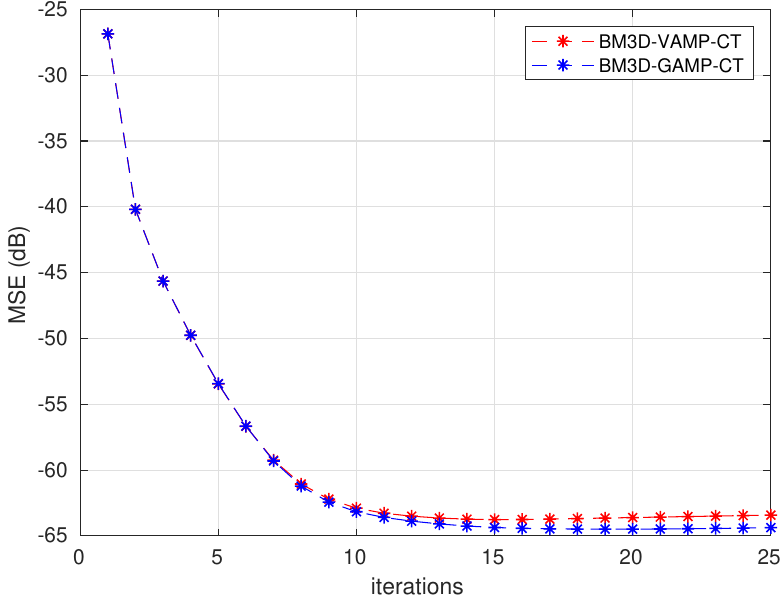}
	\end{minipage}
	\caption{MSE plot: comparison between BM3D-VAMP-CT and BM3D-GAMP-CT.}
	\label{fig:MSE_plot_GAMP_VAMP}
\end{figure}

We analyze the performance of the BM3D-GAMP-CT and BM3D-VAMP-CT which is an alternative type of message passing algorithm (as described in Section \ref{subsec:VAMP}). 
The SVD-based implementation of VAMP requires the calculation of the eigenvalues of the symmetric matrix $\*A\*A^T$, with $\*A$ being the forward system matrix. In the ASTRA toolbox, $\*A$ is defined in form of an operator, for fast computations. Therefore, it is not efficient to estimate the eigenvalues in high dimensions since it requires to generate the $M$ columns of $\*A\*A^T$ by calculating $M$ times $A(A^T(\*e_i))$, being $\*e_i$ the $i$-th unity vector, $i=1,\ldots, M$, and then calculating the eigenvalues By using the Matlab command $\mathrm{eig}$, it takes around 10 min to generate the eigenvalues. We report an estimated value for the condition number $\kappa = \frac{\lambda_{max}}{\lambda_{min}}\sim 5\times 10^4$. Moreover, Fig. \ref{fig:MSE_plot_GAMP_VAMP} shows that both algorithms behave similarly and they converge at similar values of MSE (0.8 dB difference). This simulations seem not to be in disagreement with earlier results in \cite{rangan2016}(Figs. 3 and 4) where VAMP and GAMP with damping exhibit similar MSE behavior. From the simulation, both in terms of computation and accuracy, BM3D-GAMP-CT (with preconditioning) outperforms BM3D-VAMP-CT for this CT dataset.

\section{Experimental Results}\label{sec:results}

In this section, we investigate the reconstruction quality of D-GAMP-CT on real CT data. The D-GAMP-CT framework has been applied for CT reconstruction on real luggage scans obtained using Morpho CTX5500 Air Cargo dual energy system with fan beam CT geometry. This is a single-row scanner with 476 detector channels and a 80 cm field of view. For each transversal location, two slices were acquired one at 100 KVp, the other at 198 KVp; at each energy, the full acquisition of a single slice contains 720 views/projections. The reconstruction has been performed for each energy independently and here we consider only the results obtained for 100 kVp. The reconstructed images are of $512\times 512$ array size. The total dataset contains 44 scanned slices (images).

\begin{figure}[!h]
	\centering
	\subfloat{\includegraphics[width=.49\textwidth]{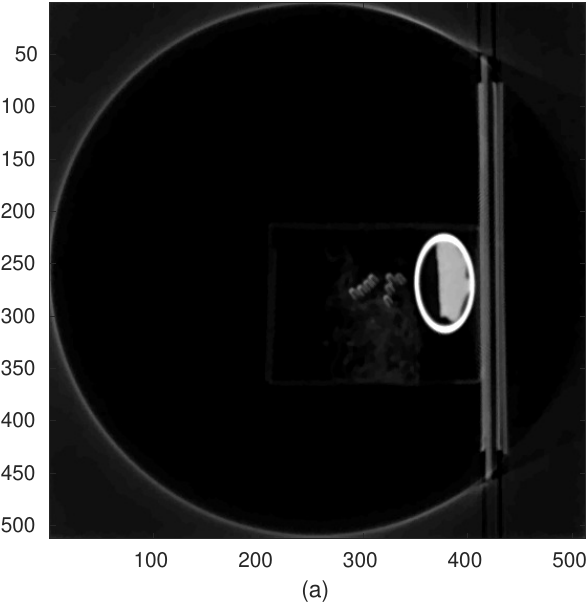}}
	\hspace{.0em}
	\subfloat{\includegraphics[width=.49\textwidth]{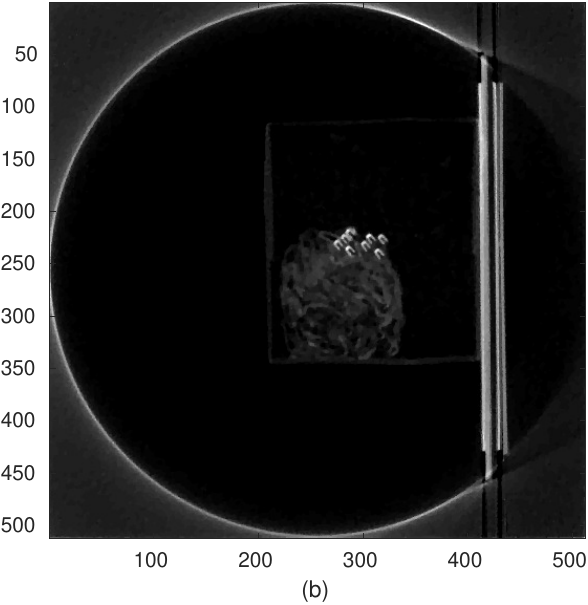}}
	\hspace{.0em}
	\subfloat{\includegraphics[width=.49\textwidth]{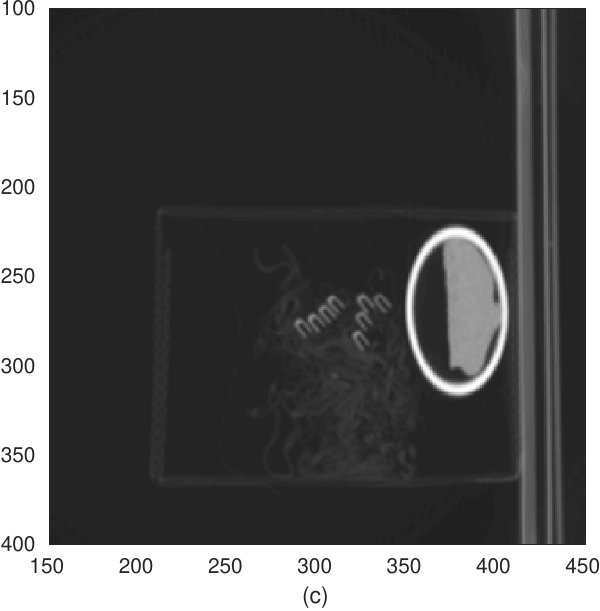}}
	\hspace{.0em}
	\subfloat{\includegraphics[width=.49\textwidth]{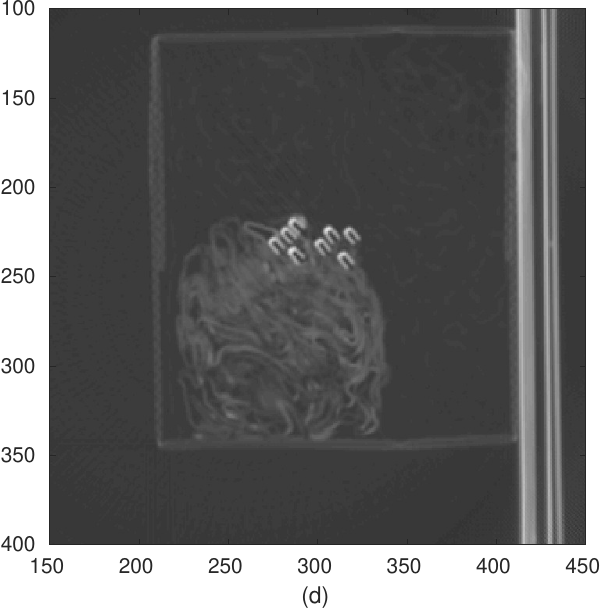}}
	\caption{Full CT reconstruction and ROI using BM3D-GAMP-CT with preconditioning and Poisson noise model: (a)-(c) slice 10, (b)-(d) slice 42.}\label{fig:imageBM3D}
\end{figure}

\begin{figure}[!h]
	\centering
	\subfloat{\includegraphics[width=.49\textwidth]{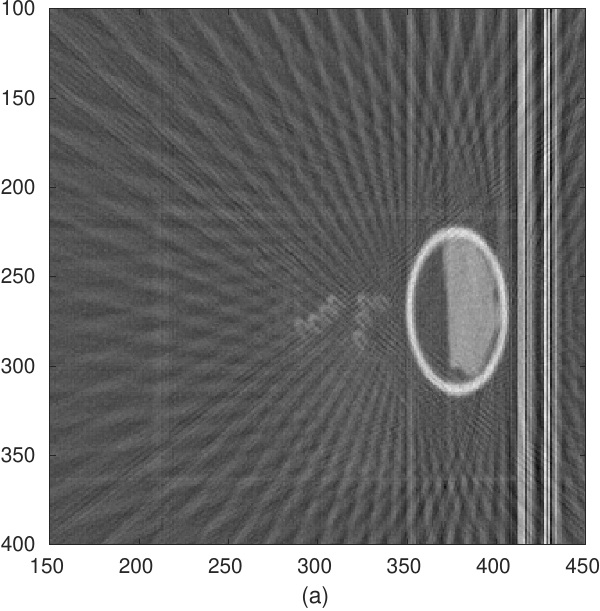}}
	\hspace{.0em}
	\subfloat{\includegraphics[width=.49\textwidth]{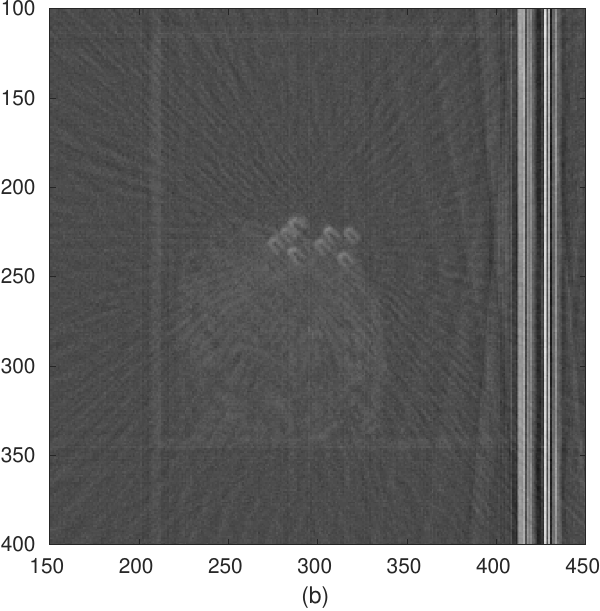}}
	\caption{Filtered Back Projection with 72 projections: (a) slice 10; (b) slice 42.}\label{fig:imageFBP}
\end{figure}

The results in Fig. \ref{fig:imageBM3D} show two slices and the zoom around the region of interest (ROI) from the reconstruction with BM3D-GAMP-CT using 72 views regularly undersampled out of the full set of views constituted of 720 views. For computing the quantitative performances in terms of PSNR and SSIM, the FBP with full number of projections (720 views) is considered as a proxy for the ground truth. In the figure it is possible to see that the scanned object contains highly resolved metal staples, bottle of fluid, wires. The number of iterations for BM3D-GAMP-CT algorithms tends to converge in around 15 iterations as shown in Section \ref{subsec:GAMP_PP}. In Figure \ref{fig:imageFBP} we show the reconstruction for 2 slices of the entire volume (10 and 42) obtained with FBP using 72 views for one of the image slices.

\subsection{Role of the Onsager term}

\begin{figure}[!h]
	\centering
	\subfloat{\includegraphics[width=.49\textwidth]{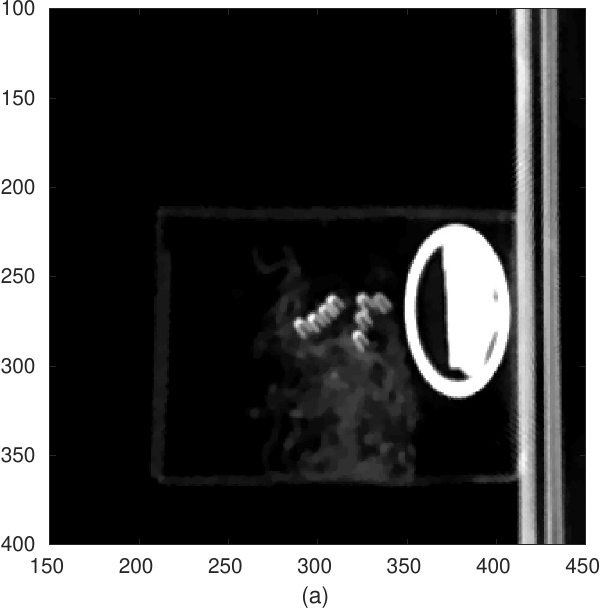}}
	\hspace{.0em}
	\subfloat{\includegraphics[width=.49\textwidth]{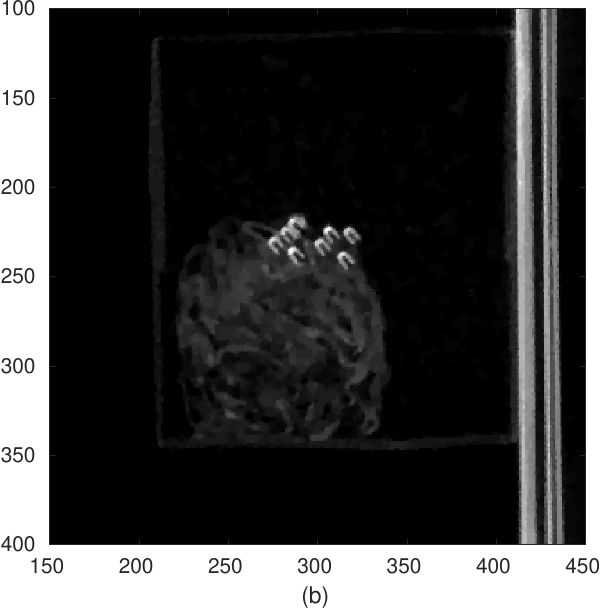}}
	\caption{CT image reconstruction using BM3D-AMP without Onsager term for: (a) slice 10; (b) slice 42.}\label{fig:imageBM3D_On}
\end{figure}

Given the similarity between iterative shrinkage algorithms and the GAMP family of algorithms, it is interesting to evaluate the importance of the Onsager term $\tau_{p}^t \*s^{t-1}$ in the D-GAMP-CT algorithm \ref{Algo:GAMP} to check whether it improves the reconstruction. Without the Onsager term the GAMP algorithm behaves like a denoising iterative thresholding algorithm \cite{metzler2014}. The reconstruction for slices 10 and 42 without the Onsager term is shown in Fig. \ref{fig:imageBM3D_On} which highlights a substantial reduction in performance in both cases, as it is also quantitatively confirmed by the PSNR value in table \ref{tab:PSNRprecon}.  

The Onsager term yields a PSNR improvement of 6 dB, for this particular CT reconstruction instance. Furthermore if we consider the algorithm with Onsager term, although the time per iteration is almost doubled because of the computation of an additional denoiser, the total running time is lower compared to the version without Onsager term because it converges in few iterations (15 respect to 40).

\begin{table}[!h]
	\centering%
	\caption{PSNR for of BM3D-GAMP-CT with/without Onsager term}\label{tab:PSNRprecon}
	\vskip 0.25cm
	\begin{tabular}{lcccc}
	\textbf{Algorithms}				&	\textbf{PSNR} [dB]	 & \textbf{SSIM} & \textbf{Time}\\
	\hline
	\hline     
	BM3D-GAMP-CT 						&	61.2 & 0.85    & 3.5 min		\\
	 BM3D-GAMP-CT without Onsager term		&	55.1 &  0.71    & 4 min\\
	 	 BM3D-GAMP with damping		&	50.3 &  0.65    & 4.5 min\\
	\hline
	\hline
	\end{tabular}
\end{table}

\subsection{Role of the Preconditioner}

\begin{figure}[!h]
	\centering
	\subfloat{\includegraphics[width=.49\textwidth]{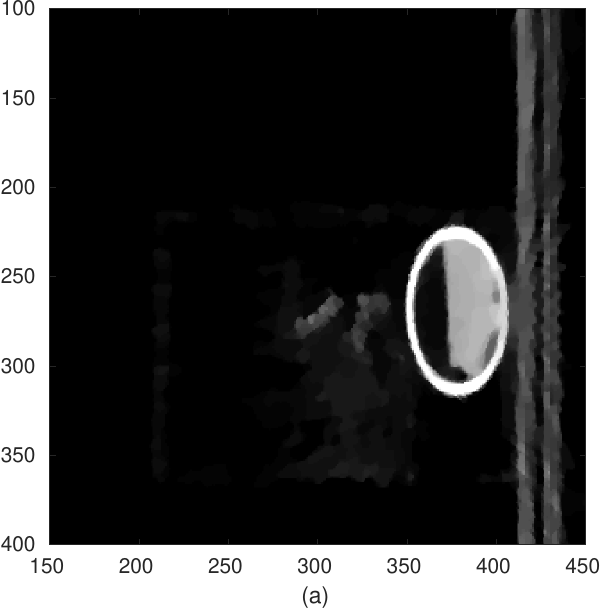}}
	\hspace{.0em}
	\subfloat{\includegraphics[width=.49\textwidth]{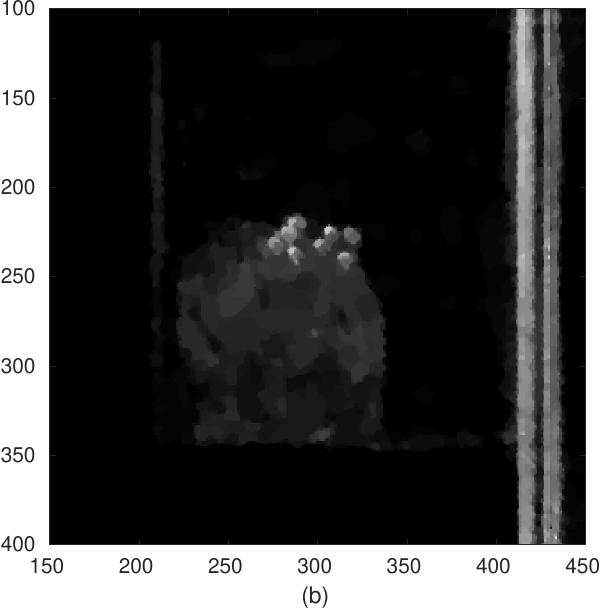}}
	\caption{BM3D-GAMP-CT reconstruction with damping $\eta_x=0.65$, $\eta_z=0.9$ for: (a) slice 10; (b) slice 42.}\label{fig:damping_Morpho}
\end{figure}

By applying the Denoising Generalized Approximate Message Passing algorithm, which includes the exponential Poisson noise model, the iteration estimates diverge. Therefore, for comparison we applied damping \cite{schniter2015,vila2015} in estimating vectorised image at iteration $t$, i.e. $\*x_t = \eta_x\*x^t  + (1-\eta_x)\*x^{t-1}$, with $0 < \eta_x < 1$ the damping factor and on the estimated linear measurement vector, i.e. $\*s^t = \eta_r\*s^t  + (1-\eta_r)\*s^{t-1}$, with $0 < \eta_r < 1$, where the vector variables $\*x$ and $\*s$ are calculated as in Algorithm \ref{Algo:GAMP}. In this case, the BM3D-GAMP-CT algorithm (without preconditioning) starts to converge for $\eta_r = 0.95$ and with a good amount of damping on the estimated $\*x^t$, $\eta_x = 0.65$. Fig. \ref{fig:damping_Morpho} shows the 
reconstruction of damped BM3D-GAMP-CT and Table \ref{tab:PSNRprecon} reports the quantitative metrics. BM3D-GAMP-CT exhibits improved PSNR and SSIM over both BM3D-GAMP-CT without the Onsager term or with damping. Furthermore, the computational time both without Onsager term or with damping is higher since the algorithms require more iterations, 40 and 25 respectively, to converge compared to 15 iterations needed by BM3D-GAMP-CT. Regarding the behavior of the damping for ill-conditioned matrices, our results are coherent with previous simulations reported in \cite{vila2015}(Fig. 1D) where the adaptive damping in GAMP does not prevent divergence for high values of condition number of the system matrix $A$. Therefore, damping is effective for low condition number (less than 10), while our Radon transform has condition number of order $\sim 10^4$. Therefore, our results confirm that damping is not very effective for high ill-conditioned matrix.

\subsection{Comparison against Plug-and-play approach}\label{subsec:GAMP_PP}

We present a comparison between D-GAMP-CT and PnP, using ADMM for solving the NLL minimization (\ref{eq:costPois}), using BM3D and TV denoisers. We are comparing 3 different methods: BM3D-GAMP-CT and the Plug-and-play approach with NLL data fidelity term BM3D-ADMM-NLL and TV-ADMM-NLL. Since the denoisers implicitly imposed in the cost function that BM3D-ADMM-NLL minimizes is different from the one of TV-ADMM-NLL, the 2 solvers yield different accuracies.

\begin{figure}[!t]
	\centering
	\subfloat{\includegraphics[width=.48\textwidth]{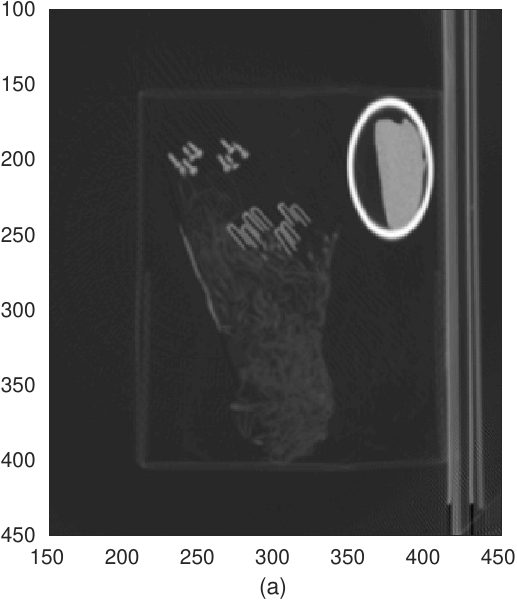}}
	\hspace{.0em}
	\subfloat{\includegraphics[width=.48\textwidth]{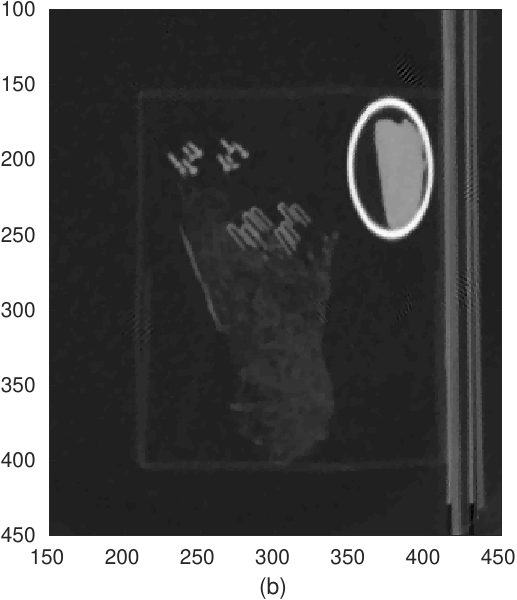}}
	\hspace{.0em}
	\subfloat{\includegraphics[width=.48\textwidth]{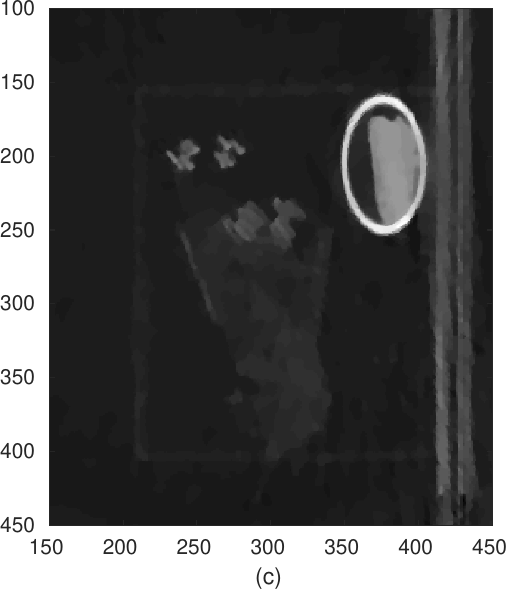}}\\
	\caption{Recontruction of slice 20 with: (a) BM3D-GAMP-CT, (b) BM3D-ADMM-NLL (PnP) (c) TV-ADMM-NLL (PnP) - NLL objective (\ref{eq:costPois}) with TV denoiser.}\label{fig:TV}
\end{figure}

\begin{figure}[!h]
	\centering
	\subfloat{\includegraphics[width=.49\textwidth]{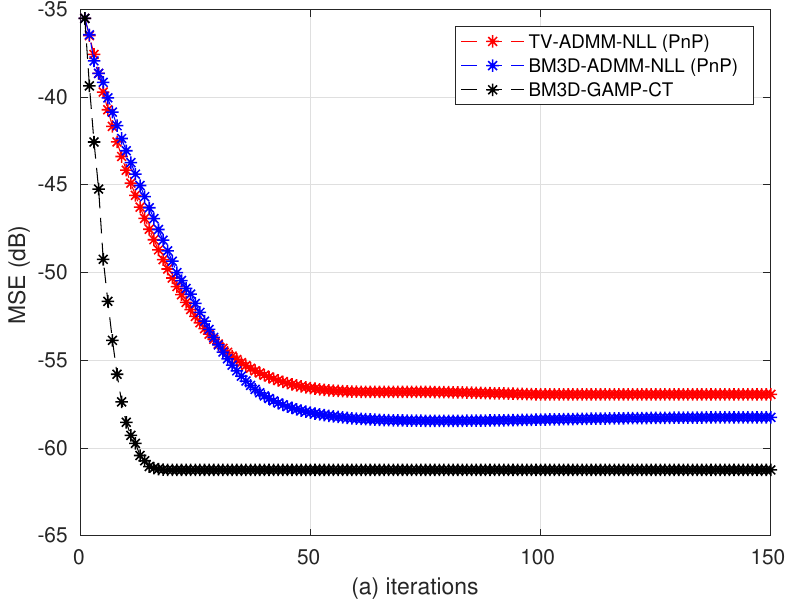}}
	\hspace{.0em}
	\subfloat{\includegraphics[width=.49\textwidth]{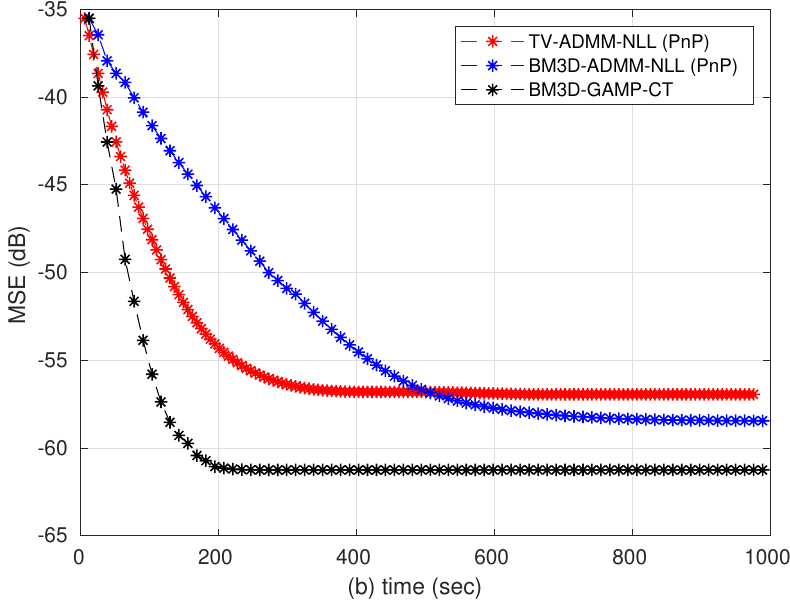}}
	\caption{Comparison between BM3D-GAMP-CT, BM3D-ADMM-NLL and TV-ADMM-NLL: (a) MSE vs iterations; (b) MSE vs running time (sec).}\label{fig:MSE_plot_GAMP_PP_GE}
\end{figure}

In Fig. \ref{fig:TV}, we show the qualitative results of one reconstructed slice. From inspection, it is clear that the reconstruction with BM3D-GAMP-CT better retains the details in the ROI. This is confirmed quantitatively in Figs. \ref{fig:MSE_plot_GAMP_PP_GE}(a)-(b) which show the convergence of the MSE against iterations and running time. Table \ref{tab:TVconf} contains the quantitative details in terms of PSNR, SSIM and time. 
If we analyze the algorithms using the same denoiser (BM3D), Fig. \ref{fig:MSE_plot_GAMP_PP_GE}(a) shows that the BM3D-GAMP-CT (line in black) converges faster both in terms of number of iterations and total running time compared to BM3D-ADMM-NLL for Plug-and-Play optimization. Furthermore, at convergence BM3D-GAMP-CT yields to a reduction in PSNR of 3 dB.

\begin{table}[!h]
	\centering%
	\caption{PSNR of BM3D-GAMP-CT, BM3D-ADMM-NLL  and TV-ADMM-NLL}\label{tab:TVconf}
	\vskip 0.25cm
		\begin{tabular}{lcccc}
			\textbf{Algorithms}				&	\textbf{PSNR} [dB] & \textbf{SSIM}  &  \textbf{Time}\\
			\hline
			\hline       
			BM3D-GAMP-CT 						&	61.3 &  0.85   & 3.5 min		\\
			BM3D-ADMM-NLL (PnP)						&	58.2 &  0.78    & 11 min\\ 
			TV-ADMM-NLL	(PnP)					&	56.8 &  0.72    & 6.5 min\\
			\hline
			\hline
		\end{tabular}
\end{table}

We have tried a different denoiser, TV, for the PP framework and we can observe from Figs. \ref{fig:MSE_plot_GAMP_PP_GE} that at earlier iterations BM3D-ADMM-NLL and TV-ADMM-NLL perform similarly because the input estimate is highly noisy but at convergence the BM3D reaches higher PSNR and SSIM. Interestingly, the running time of TV-ADMM-NLL is lower compared to BM3D-ADMM-NLL since they converge almost at the same number of iterations but the TV denoiser is computationally faster than BM3D. However, BM3D-GAMP-CT is faster in total running time than TV-ADMM-NLL because BM3D-GAMP-CT converges in almost $\frac{1}{6}$ the number of iterations of TV-ADMM-NLL.

\subsection{State Evolution Analysis}\label{sec:SE}

An important aspect of D-GAMP-CT is its internal variance estimate within the algorithm and in the SE equations. This not only provides an estimate of uncertainty with the algorithm, it also essentially allows the algorithm to adapt its "step size" on the fly \cite{rangan2015}. 
It is therefore instructive to see how precise such an estimate is. If accurate, this term should ensure a fast convergence rate of the algorithm. Given the actual MSE estimate, taking as reference the full views FBP reconstruction, we can calculate the predicted MSE at the next iteration and compare with the actual estimate. 

\begin{figure}[!h]
	\begin{minipage}[!h]{\linewidth}
		\centering
		\includegraphics[width=.55\textwidth]{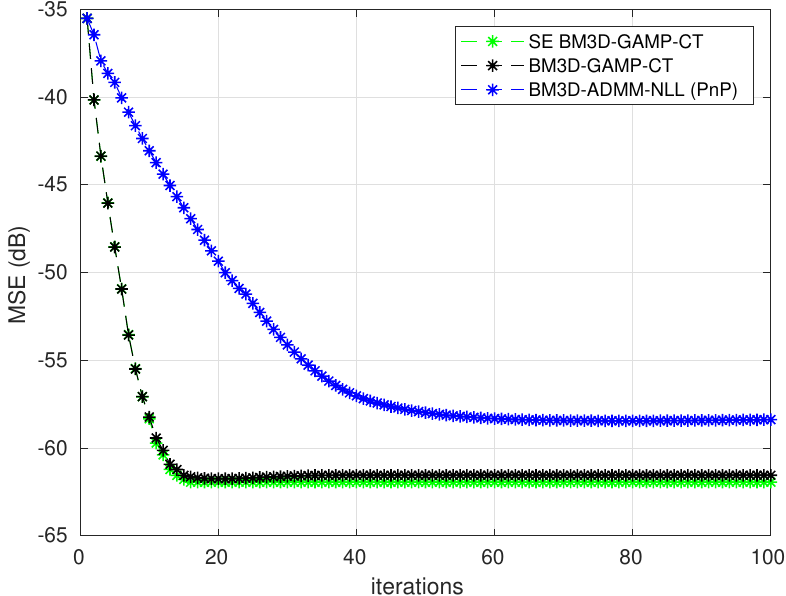} 
	\end{minipage}
	\caption{Deterministic state evolution and MSE estimates using BM3D-AMP-CT and BM3D-ADMM-NLL.}\label{fig:MSE_SE}
\end{figure}

Our empirical results run on the experimental data acquired using the 2D fan beam CT geometry show that the SE prediction gives an accurate estimate of the true MSE of BM3D-GAMP-CT at each iteration as depicted in Figure \ref{fig:MSE_SE}.

Fig. \ref{fig:MSE_SE} shows the theoretical SE (prediction of the MSE) of the proposed D-GAMP-CT algorithm (in red) and the actual performance of the algorithm with BM3D. The 2 curves are almost matched; the little discrepancy is due to the fact that the SE is derived under the "matched" case indicated in Eq. (\ref{eq:SE_matches}), i.e. exact estimation of the variance of the MMSE for the Poisson noise while, as described in Eq. (\ref{eq:post_exp}), we estimate it by using the Laplace approximation method. Moreover, this plot highlights that the Laplace approximation is accurate since the error with the SE is only around 0.1 dB. Furthermore, the Plug-and-play solver BM3D-ADMM-NLL (black curve) performs always worse than D-GAMP-CT and also the MSE behavior of BM3D-ADMM-NLL cannot be predicted by SE equations.

\section{Conclusions and Future Research}
In this work, we have presented a Generalized Approximate Message Passing type of iterative algorithm for solving X-ray CT reconstruction from a limited number of projections. The proposed framework relies on the design of an appropriate preconditioner for the ill-conditioned measurement matrix and a statistical model for the non linear Poisson measurement noise. In addition, exploiting the flexibility of the GAMP framework we can decouple the action of the preconditioner from the noise model, which is not possible with optimization solvers for minimizing the Plug-and-play PP-WLS objective function.

We have experimentally shown the important role of the Onsager term regarding reconstruction performance improvement and the ability of the state evolution analysis to estimate the current MSE through the iterations. 
Numerical results on simulations and experimental Cargo dataset demonstrate how the D-GAMP-CT framework provides high reconstruction accuracy and reduced running time compared to state-of-the-art Plug-and-play optimization iterative algorithms for CT reconstruction. In addition D-GAMP-CT allows different prior image models to be used on the signal by employing different denoisers. 

Finally, further acceleration of the D-GAMP-CT may be possible utilizing the Ordered Subsets principle \cite{erdogan1999}, however its implementation is not straight forward within the GAMP framework and is also left for future research.

\section*{Acknowledgment}
\addcontentsline{toc}{section}{Acknowledgment}
The authors are grateful to Jonathan H. Mason from The University of Edinburgh for providing invaluable comments and for improving the quality of this work with insightful discussions. The authors thank Walter Garms, Morpho Detection Inc., for providing real CT datasets and the anonymous reviewers whose comments helped improving the manuscript. This work was supported by U.S. Department of Homeland Security, Science and Technology Directorate, Explosives Division, BAA 13-05, Contract \# HSHQDC-14-C-B0048, EPSRC platform grant EP/J015180/1 and the ERC Advanced grant, project C-SENSE, (ERC-ADG-2015-694888). The research leading to these results has received funding from the European Union's Horizon 2020 research and innovation programme under the Marie Sklodowska-Curie grant agreement no. 713683 (COFUNDfellowsDTU).

\appendix

\section{Laplace method for approximating the posterior mean of the Nonlinear noise distribution}\label{sec_app:Laplace}

In order to evaluate the expression in (\ref{eq:post_exp}), we write the ration of integrals in the following form (where we have not indicated the iteration $t$ for notation simplicity)
\begin{equation}
\Expect(z_a|p_a,y_a,\tau_p)=\frac{\int_{\mathbb{R}_{\geq 0}} g(z_a)e^{\log p(y_a|z_a)} \pi(z_a)dz_a}{\int_{\mathbb{R}_{\geq 0}} e^{\log p(y_a|z_a)} \pi(z_a)dz_a}    
\end{equation}

\noindent where $\pi(z_a) = e^{-\frac{1}{2 \tau_p} (z_a-\hat{p_a})^2}$ and $g(z_a) = z_a$. We set
\begin{eqnarray}
L &=& \log\pi + \frac{1}{M}\log p(y_a|z_a) = -\frac{1}{2 \tau_p} (z_a - p_a)^2 + \frac{1}{M}\bigg[-z_a y_a-e^{-z_a} - \log(y_a!)\bigg]\nonumber\\
L^* &=& \log z_a + L = \log g(z_a) + \log\pi(z_a) + \frac{1}{M}\log p(y_a|z_a)\nonumber\\
 &=& \log z_a -\frac{1}{2 \tau_p} (z_a - p_a)^2  + \frac{1}{M}\bigg[-z_a y_a-e^{-z_a} - \log(y_a!)\bigg]
\end{eqnarray}

\noindent Therefore, the MMSE can be written as 
\begin{equation}\label{eq:post_exp2} \Expect(z_a|p_a,y_a,\tau_p) = \frac{\int_{\mathbb{R}_{\geq 0}} e^{M\cdot L^*} dz_a}
{\int_{\mathbb{R}_{\geq 0}} e^{M\cdot L} dz_a} 
\end{equation}

\noindent We consider the probability density function $L(z_a)$ which we expect to have a peak at the point $z_{0_a}$ and the Taylor-expansion of $L(z_a)$ at $z_{0_a}$ is 
\begin{equation}
L(z_a) \approx L(z_{0_a}) - \frac{1}{2}\frac{\partial^2L(z_a)}{\partial z_a^2}(z_a - z_{0_a})^2 + \ldots
\end{equation}

\noindent The Laplace's method \cite{tierney1986} is a way to approximate $L(z_a)$ by an unnormalized Gaussian and approximate the partition function $Z_P = \int L(z_a)dz_a$ with the one of the Gaussian 
\begin{equation}
Z_Q = L(z_{0_a})\sqrt{2\pi} 
\end{equation}

\noindent The Laplace approximation leads to 
\begin{eqnarray}
\int e^{m L(z_a)}dz_a &\approx& \int e^{mL(z_{0_a}) - m(z_a - z_{0_a})^2/(2\sigma^2)}dz_a \nonumber\\
&=& \sqrt{2\pi}\sigma n^{-1/2}e^{mL(z_{0_a})}
\end{eqnarray}
with $\sigma^2= -\frac{1}{L^{''}(z_{0_a})}$;  this integral form is similar to the one in Eq. (\ref{eq:post_exp}) for the numerator and denominator respectively. Considering the denominator, we need to calculate the points where the derivative is zero in order to find $z_{0_a}$ :
\begin{eqnarray*}
\frac{\partial L(z_a)}{\partial z_a} & = & -\frac{1}{\tau_p}(z_a - p_a) - y_a + e^{-z_a}
\end{eqnarray*}

\noindent with $y_a\in\mathbb{Z}_+$, $ z_a = \left[\tilde{\bm\Phi}\*x\right]_a\in\mathbb{R}_{\geq 0}$; then, to find $\frac{\partial L(z_a)}{\partial z_a} = 0$, it results
\begin{eqnarray*}
-\frac{(z_a - p_a)}{\tau_p} - y_a + e^{-z_a} &=& 0\\
\log\bigg[-\frac{(z_a - p_a)}{\tau_p} - y_a\bigg] &=& z_a
\end{eqnarray*}

\noindent Finally, the second derivative is
\begin{eqnarray*}
\frac{\partial^2 L(z_a)}{\partial z_a^2}\bigg|_{z_{a_0}} & = & -\frac{z_{a_0}}{\tau^p} - e^{-z_{a_0}}
\end{eqnarray*}

\noindent Similar procedure for the numerator ($\sigma^*$ and $z_{a_0}^*$); therefore, taking the ratio of the 2 approximations it yields to 
\begin{equation}\label{appr}
\mathbb{E}[z_a|{p_a}] = \frac{\sigma^*}{\sigma}e^{L^*(z_{a_0}^*)-L(z_{a_0})}
\end{equation}

\noindent For the variance, given the approximation \ref{appr}, we can use the standard formula

\begin{equation}
\mathrm{Var}[z_a|p_a] = \mathbb{E}[z_a^2|{p_a}] - \mathbb{E}[z_a|{p_a}]^2
\end{equation}


\bibliographystyle{siamplain}
\bibliography{references}

\begin{thebibliography}{10}

\bibitem{averbuch2008first}
{\sc A.~Averbuch, R.~Coifman, D.~Donoho, M.~Israeli, and Y.~Shkolnisky}, {\em A
  framework for discrete integral transformations {I} - the pseudopolar fourier
  transform}, SIAM Journal on Scientific Computing, 30 (2008), pp.~764--784.

\bibitem{averbuch2001}
{\sc A.~Averbuch, R.~Coifman, D.~Donoho, M.~Israeli, and J.~Walden}, {\em Fast
  slant stack: A notion of radon transform for data in a cartesian grid which
  is rapidly computible, algebraically exact, geometrically faithful and
  invertible}, SIAM Scientific Computing,  (2001).

\bibitem{averbuch2008}
{\sc A.~Averbuch, R.~Coifman, D.~L. Donoho, M.~Israeli, Y.~Shkolnisky, and
  I.~Sedelnikov}, {\em A framework for discrete integral transformations {II} -
  the 2{D} discrete radon transform}, SIAM Journal on Scientific Computing, 30
  (2008), pp.~785--803.

\bibitem{bayati2011}
{\sc M.~Bayati and A.~Montanari}, {\em The dynamics of message passing on dense
  graphs, with applications to compressed sensing}, IEEE Transactions on
  Information Theory, 57 (2011), pp.~764--785.

\bibitem{bayati2011lasso}
{\sc M.~Bayati and A.~Montanari}, {\em The lasso risk for gaussian matrices},
  IEEE Transactions on Information Theory, 58 (2011), pp.~1997--2017.

\bibitem{beister2012}
{\sc M.~Beister, D.~Kolditz, and W.~A. Kalender}, {\em Iterative reconstruction
  methods in {X}-ray {CT}}, Physica medica, 28 (2012), pp.~94--108.

\bibitem{berthier2020}
{\sc R.~Berthier, A.~Montanari, and P.-M. Nguyen}, {\em State evolution for
  approximate message passing with non-separable functions}, Information and
  Inference: A Journal of the IMA, 9 (2020), pp.~33--79.

\bibitem{bruyant2000}
{\sc P.~Bruyant, J.~Sau, and J.~Mallet}, {\em Streak artifact reduction in
  filtered backprojection using a level line--based interpolation method},
  Journal of Nuclear Medicine, 41 (2000), pp.~1913--1919.

\bibitem{buzzard2018}
{\sc G.~T. Buzzard, S.~H. Chan, S.~Sreehari, and C.~A. Bouman}, {\em
  Plug-and-play unplugged: Optimization-free reconstruction using consensus
  equilibrium}, SIAM Journal on Imaging Sciences, 11 (2018), pp.~2001--2020.

\bibitem{cakmak2014}
{\sc B.~Cakmak, O.~Winther, and B.~H. Fleury}, {\em {S-AMP}: Approximate
  message passing for general matrix ensembles}, in IEEE Information Theory
  Workshop (ITW), 2014, pp.~192--196.

\bibitem{calatroni2017}
{\sc L.~Calatroni and A.~Chambolle}, {\em Backtracking strategies for
  accelerated descent methods with smooth composite objectives}, arXiv preprint
  arXiv:1709.09004,  (2017).

\bibitem{chun2014}
{\sc S.~Y. Chun, Y.~Dewaraja, and J.~A. Fessler}, {\em Alternating direction
  method of multiplier for tomography with nonlocal regularizers}, IEEE
  Transactions on Medical Imaging, 33 (2014), pp.~1960--1968.

\bibitem{Clinthorne1993}
{\sc N.~H. Clinthorne, T.~S. Pan, P.~C. Chiao, W.~L. Rogers, and J.~A. Stamos},
  {\em Preconditioning methods for improved convergence rates in iterative
  reconstructions}, IEEE Transactions on Medical Imaging, 12 (1993),
  pp.~78--83.

\bibitem{dabov2007}
{\sc K.~Dabov, A.~Foi, V.~Katkovnik, and K.~Egiazarian}, {\em Image denoising
  by sparse 3-d transform-domain collaborative filtering}, IEEE Transactions on
  Image Processing, 16 (2007), pp.~2080--2095.

\bibitem{danielyan2012}
{\sc A.~Danielyan, V.~Katkovnik, and K.~Egiazarian}, {\em {BM3D} frames and
  variational image deblurring}, IEEE Transactions on Image Processing, 21
  (2012), pp.~1715--1728.

\bibitem{ding2018}
{\sc Q.~Ding, Y.~Long, X.~Zhang, and J.~A. Fessler}, {\em Statistical image
  reconstruction using mixed poisson-gaussian noise model for x-ray ct}, arXiv
  preprint arXiv:1801.09533,  (2018).

\bibitem{donoho2009}
{\sc D.~Donoho, A.~Maleki, and A.~Montanari}, {\em Message-passing algorithms
  for compressed sensing}, Proceedings of the National Academy of Sciences, 106
  (2009), pp.~18914--18919.

\bibitem{elbakri2002}
{\sc I.~A. Elbakri and J.~A. Fessler}, {\em Statistical image reconstruction
  for polyenergetic {X}-ray computed tomography}, IEEE Transactions on Medical
  Imaging, 21 (2002), pp.~89--99.

\bibitem{erdogan1999}
{\sc H.~Erdogan and J.~A. Fessler}, {\em Ordered subsets algorithms for
  transmission tomography}, Physics in medicine and biology, 44 (1999),
  p.~2835.

\bibitem{MetzlerMatlab}
{\sc C.~M. et~al.}, {\em {D-AMP} toolbox}.
\newblock \url{https://github.com/ricedsp/D-AMP_Toolbox}, 2017.

\bibitem{RanganMatlab}
{\sc S.~R. et~al.}, {\em Gampmatlab toolbox}.
\newblock
  \url{http://gampmatlab.wikia.com/wiki/Generalized_Approximate_Message_Passing},
  2017.

\bibitem{fessler1999}
{\sc J.~A. Fessler and S.~Booth}, {\em Conjugate-gradient preconditioning
  methods for shift-variant pet image reconstruction}, IEEE Transactions on
  Image Processing, 8 (1999), pp.~688--699.

\bibitem{fessler2003nonuniform}
{\sc J.~A. Fessler and B.~P. Sutton}, {\em Nonuniform fast fourier transforms
  using min-max interpolation}, IEEE transactions on signal processing, 51
  (2003), pp.~560--574.

\bibitem{Fletcher2018}
{\sc A.~K. Fletcher, S.~Rangan, S.~Sarkar, and P.~Schniter}, {\em Plug-in
  estimation in high-dimensional linear inverse problems: A rigorous analysis},
  in NIPS, 2018.

\bibitem{florea2017}
{\sc M.~I. Florea and S.~A. Vorobyov}, {\em A generalized accelerated composite
  gradient method: Uniting nesterov's fast gradient method and fista}, arXiv
  preprint arXiv:1705.10266,  (2017).

\bibitem{gonzales2014}
{\sc B.~Gonzales, D.~Spronk, Y.~Cheng, A.~Tucker, M.~Beckman, O.~Zhou, and
  J.~Lu}, {\em Rectangular fixed-gantry {CT} prototype: combining cnt {X}-ray
  sources and accelerated compressed sensing-based reconstruction}, IEEE
  Access, 2 (2014), pp.~971--981.

\bibitem{guo2015}
{\sc C.~Guo and M.~E. Davies}, {\em Near optimal compressed sensing without
  priors: Parametric sure approximate message passing.}, IEEE Trans. Signal
  Processing, 63 (2015), pp.~2130--2141.

\bibitem{huang2013}
{\sc J.~Huang, Y.~Zhang, J.~Ma, D.~Zeng, Z.~Bian, S.~Niu, Q.~Feng, Z.~Liang,
  and W.~Chen}, {\em Iterative image reconstruction for sparse-view {CT} using
  normal-dose image induced total variation prior}, PloS one, 8 (2013),
  p.~e79709.

\bibitem{jorgensen2011}
{\sc J.~H. J{\o}rgensen, T.~L. Jensen, P.~C. Hansen, S.~H. Jensen, E.~Y. Sidky,
  and X.~Pan}, {\em Accelerated gradient methods for total-variation-based {CT}
  image reconstruction}, arXiv:1105.4002,  (2011).

\bibitem{kamilov2017plug}
{\sc U.~S. Kamilov, H.~Mansour, and B.~Wohlberg}, {\em A plug-and-play priors
  approach for solving nonlinear imaging inverse problems}, IEEE Signal
  Processing Letters, 24 (2017), pp.~1872--1876.

\bibitem{kim2015}
{\sc D.~Kim, S.~Ramani, and J.~A. Fessler}, {\em Combining ordered subsets and
  momentum for accelerated {X}-ray {CT} image reconstruction}, IEEE
  Transactions on Medical Imaging, 34 (2015), pp.~167--178.

\bibitem{ma2017}
{\sc J.~Ma and L.~Ping}, {\em Orthogonal amp}, IEEE Access, 5 (2017),
  pp.~2020--2033.

\bibitem{manoel2015}
{\sc A.~Manoel, F.~Krzakala, E.~Tramel, and L.~Zdeborova}, {\em Swept
  approximate message passing for sparse estimation}, in International
  Conference on Machine Learning, 2015, pp.~1123--1132.

\bibitem{matej2004iterative}
{\sc S.~Matej, J.~A. Fessler, and I.~Kazantsev}, {\em Iterative tomographic
  image reconstruction using fourier-based forward and back-projectors}, IEEE
  transactions on medical imaging, 23 (2004), pp.~401--412.

\bibitem{metzler2014}
{\sc C.~A. Metzler, A.~Maleki, and R.~G. Baraniuk}, {\em From denoising to
  compressed sensing}, IEEE Transactions on Information Theory, 62 (2016),
  pp.~5117--5144.

\bibitem{nagy1996}
{\sc J.~G. Nagy, R.~J. Plemmons, and T.~C. Torgersen}, {\em Iterative image
  restoration using approximate inverse preconditioning}, IEEE Transactions on
  Image Processing, 5 (1996), pp.~1151--1162.

\bibitem{nilchian2013}
{\sc M.~Nilchian, C.~Vonesch, P.~Modregger, M.~Stampanoni, and M.~Unser}, {\em
  Fast iterative reconstruction of differential phase contrast {X}-ray
  tomograms}, Optics express, 21 (2013), pp.~5511--5528.

\bibitem{niu2014}
{\sc S.~Niu, Y.~Gao, Z.~Bian, J.~Huang, W.~Chen, G.~Yu, Z.~Liang, and J.~Ma},
  {\em Sparse-view {X}-ray {CT} reconstruction via total generalized variation
  regularization}, Physics in medicine and biology, 59 (2014), p.~2997.

\bibitem{nuyts2013}
{\sc J.~Nuyts, B.~De~Man, J.~A. Fessler, W.~Zbijewski, and F.~J. Beekman}, {\em
  Modelling the physics in the iterative reconstruction for transmission
  computed tomography}, Physics in medicine and biology, 58 (2013), p.~R63.

\bibitem{o2006fourier}
{\sc Y.~O'Connor and J.~A. Fessler}, {\em Fourier-based forward and
  back-projectors in iterative fan-beam tomographic image reconstruction}, IEEE
  transactions on medical imaging, 25 (2006), pp.~582--589.

\bibitem{ono2017primal}
{\sc S.~Ono}, {\em Primal-dual plug-and-play image restoration}, IEEE Signal
  Processing Letters, 24 (2017), pp.~1108--1112.

\bibitem{opper2005}
{\sc M.~Opper and O.~Winther}, {\em Expectation consistent approximate
  inference}, Journal of Machine Learning Research, 6 (2005), pp.~2177--2204.

\bibitem{o2015}
{\sc B.~O’donoghue and E.~Candes}, {\em Adaptive restart for accelerated
  gradient schemes}, Foundations of computational mathematics, 15 (2015),
  pp.~715--732.

\bibitem{perelli2015compressive}
{\sc A.~Perelli and M.~E. Davies}, {\em Compressive computed tomography image
  reconstruction with denoising message passing algorithms}, in Signal
  Processing Conference (EUSIPCO), 2015 23rd European, IEEE, 2015,
  pp.~2806--2810.

\bibitem{potts2000new}
{\sc D.~Potts and G.~Steidl}, {\em New fourier reconstruction algorithms for
  computerized tomography}, in Wavelet Applications in Signal and Image
  Processing VIII, vol.~4119, International Society for Optics and Photonics,
  2000, pp.~13--24.

\bibitem{ramani2008}
{\sc S.~Ramani, T.~Blu, and M.~Unser}, {\em Monte-carlo {SURE}: A black-box
  optimization of regularization parameters for general denoising algorithms},
  IEEE Transactions on Image Processing, 17 (2008), pp.~1540--1554.

\bibitem{ramani2012}
{\sc S.~Ramani and J.~A. Fessler}, {\em A splitting-based iterative algorithm
  for accelerated statistical {X}-ray {CT} reconstruction}, IEEE Transactions
  on Medical Imaging, 31 (2012), pp.~677--688.

\bibitem{ramani2016}
{\sc S.~Ramani, X.~Wang, L.~Fu, and M.~Lexa}, {\em Denoising-based accelerated
  statistical iterative reconstruction for {X}-ray {CT}}, in 4th International
  Meeting on Image Formation in X-Ray Computed Tomography, 2016, pp.~337--340.

\bibitem{rangan2010}
{\sc S.~Rangan}, {\em Generalized approximate message passing for estimation
  with random linear mixing}, arXiv preprint arXiv:1010.5141,  (2010).

\bibitem{rangan2011}
{\sc S.~Rangan}, {\em Generalized approximate message passing for estimation
  with random linear mixing}, in IEEE International Symposium on Information
  Theory Proceedings, 2011, pp.~2168--2172.

\bibitem{rangan2015}
{\sc S.~Rangan, A.~K. Fletcher, P.~Schniter, and U.~S. Kamilov}, {\em Inference
  for generalized linear models via alternating directions and {B}ethe free
  energy minimization}, IEEE Transactions on Information Theory, 63 (2016),
  pp.~676--697.

\bibitem{rangan2016}
{\sc S.~Rangan, P.~Schniter, and A.~K. Fletcher}, {\em Vector approximate
  message passing}, IEEE Transactions on Information Theory, 65 (2019),
  pp.~6664--6684.

\bibitem{rangan2014}
{\sc S.~Rangan, P.~Schniter, A.~K. Fletcher, and S.~Sarkar}, {\em On the
  convergence of approximate message passing with arbitrary matrices}, IEEE
  Transactions on Information Theory, 65 (2019), pp.~5339--5351.

\bibitem{reehorst2018regularization}
{\sc E.~T. Reehorst and P.~Schniter}, {\em Regularization by denoising:
  Clarifications and new interpretations}, IEEE Transactions on Computational
  Imaging, 5 (2018), pp.~52--67.

\bibitem{romano2017little}
{\sc Y.~Romano, M.~Elad, and P.~Milanfar}, {\em The little engine that could:
  Regularization by denoising (red)}, SIAM Journal on Imaging Sciences, 10
  (2017), pp.~1804--1844.

\bibitem{sarkar2019}
{\sc S.~Sarkar, A.~K. Fletcher, S.~Rangan, and P.~Schniter}, {\em Bilinear
  recovery using adaptive vector-amp}, IEEE Transactions on Signal Processing,
  67 (2019), pp.~3383--3396.

\bibitem{sauer1993}
{\sc K.~Sauer and C.~Bouman}, {\em A local update strategy for iterative
  reconstruction from projections}, IEEE Transactions on Signal Processing, 41
  (1993), pp.~534--548.

\bibitem{Schniter2017}
{\sc P.~Schniter, S.~Rangan, , and A.~K. Fletcher}, {\em Plug-and-play image
  recovery using vector amp}, in International BASP Frontiers workshop 2017,
  2017, \url{http://www2.ece.ohio-state.edu/~schniter/pdf/basp17_poster.pdf}.

\bibitem{schniter2015}
{\sc P.~Schniter and S.~Rangan}, {\em Compressive phase retrieval via
  generalized approximate message passing}, IEEE Transactions on Signal
  Processing, 63 (2015), pp.~1043--1055.

\bibitem{schniter2016c}
{\sc P.~Schniter, S.~Rangan, and A.~K. Fletcher}, {\em Vector approximate
  message passing for the generalized linear model}, in 50th Asilomar
  Conference on Signals, Systems and Computers, 2016, pp.~1525--1529.

\bibitem{sharma2019}
{\sc M.~Sharma, S.~Metzler, C. A.and~Nagesh, O.~Cossairt, R.~G. Baraniuk, and
  A.~Veeraraghavan}, {\em Inverse scattering via transmission matrices:
  Broadband illumination and fast phase retrieval algorithms}, IEEE
  Transactions on Computational Imaging,  (2019).

\bibitem{sidky2006}
{\sc E.~Y. Sidky, C.-M. Kao, and X.~Pan}, {\em Accurate image reconstruction
  from few-views and limited-angle data in divergent-beam ct}, Journal of X-ray
  Science and Technology, 14 (2006), pp.~119--139.

\bibitem{Sreehari2016}
{\sc S.~Sreehari, S.~V. Venkatakrishnan, and B.~Wohlberg}, {\em Plug-and-play
  priors for bright field electron tomography and sparse interpolation}, IEEE
  Transactions on Computational Imaging, 2 (2016), pp.~408--423.

\bibitem{Baron2016}
{\sc J.~Tan, Y.~Ma, H.~Rueda, D.~Baron, and G.~R. Arce}, {\em Compressive
  hyperspectral imaging via approximate message passing}, IEEE Journal of
  Selected Topics in Signal Processing, 10 (2016), pp.~389--401.

\bibitem{tierney1986}
{\sc L.~Tierney and J.~B. Kadane}, {\em Accurate approximations for posterior
  moments and marginal densities}, Journal of the American statistical
  association, 81 (1986), pp.~82--86.

\bibitem{van2016}
{\sc W.~van Aarle, W.~J. Palenstijn, J.~Cant, E.~Janssens, F.~Bleichrodt,
  A.~Dabravolski, J.~De~Beenhouwer, K.~J. Batenburg, and J.~Sijbers}, {\em Fast
  and flexible {X}-ray tomography using the astra toolbox}, Optics express, 24
  (2016), pp.~25129--25147.

\bibitem{venkatakrishnan2013}
{\sc S.~V. Venkatakrishnan, C.~A. Bouman, and B.~Wohlberg}, {\em Plug-and-play
  priors for model based reconstruction}, in 2013 IEEE Global Conference on
  Signal and Information Processing, IEEE, 2013, pp.~945--948.

\bibitem{vila2015}
{\sc J.~Vila, P.~Schniter, S.~Rangan, F.~Krzakala, and L.~Zdeborov{\'a}}, {\em
  Adaptive damping and mean removal for the generalized approximate message
  passing algorithm}, in 2015 IEEE International Conference on Acoustics,
  Speech and Signal Processing (ICASSP), 2015, pp.~2021--2025.

\bibitem{vila2013}
{\sc J.~P. Vila and P.~Schniter}, {\em Expectation-maximization
  gaussian-mixture approximate message passing}, IEEE Transactions on Signal
  Processing, 61 (2013), pp.~4658--4672.

\bibitem{vila2014empirical}
{\sc J.~P. Vila and P.~Schniter}, {\em An empirical-bayes approach to
  recovering linearly constrained non-negative sparse signals}, IEEE
  Transactions on Signal Processing, 62 (2014), pp.~4689--4703.

\bibitem{wang2004image}
{\sc Z.~Wang, A.~C. Bovik, H.~R. Sheikh, and E.~P. Simoncelli}, {\em Image
  quality assessment: from error visibility to structural similarity}, IEEE
  transactions on image processing, 13 (2004), pp.~600--612.

\bibitem{whiting2006}
{\sc B.~Whiting, P.~Massoumzadeh, O.~Earl, J.~O’Sullivan, D.~Snyder, and
  J.~Williamson}, {\em Properties of preprocessed sinogram data in {X}-ray
  computed tomography}, Medical physics, 33 (2006), pp.~3290--3303.

\bibitem{yu2011}
{\sc Z.~Yu, J.-B. Thibault, C.~Bouman, K.~Sauer, and J.~Hsieh}, {\em Fast
  model-based {X}-ray {CT} reconstruction using spatially nonhomogeneous {ICD}
  optimization}, IEEE Transactions on Image Processing, 20 (2011),
  pp.~161--175.

\bibitem{deepCNN2017}
{\sc K.~Zhang, W.~Zuo, Y.~Chen, D.~Meng, and L.~Zhang}, {\em Beyond a gaussian
  denoiser: Residual learning of deep cnn for image denoising}, IEEE
  Transactions on Image Processing, 26 (2017), pp.~3142--3155.

\end{thebibliography}

\end{document}